\documentclass[fleqn,usenatbib]{mnras}

\usepackage{newtxtext,newtxmath}
\usepackage[T1]{fontenc}
\usepackage{ae,aecompl}
\usepackage{graphicx}	% Including figure files
\usepackage{amsmath}	% Advanced maths commands
\usepackage{amssymb}	% Extra maths symbols
\usepackage{multicol}        % Multi-column entries in tables
\usepackage{bm}		% Bold maths symbols, including upright Greek
\usepackage{pdflscape}	% Landscape pages
\usepackage{soul}
\usepackage{enumitem}
\usepackage{pdflscape}	% Landscape pages
\usepackage[dvipsnames]{xcolor}
\newcommand{\dd}{\mathrm{d}}
\newcommand{\Mpc}{\mathrm{Mpc}}
\newcommand{\Myr}{\mathrm{Myr}}
\newcommand{\Gyr}{\mathrm{Gyr}}
\newcommand{\kpc}{\mathrm{kpc}}
\newcommand{\pc}{\mathrm{pc}}
\newcommand{\cmcube}{\mathrm{cm}^{-3}}
\newcommand{\protonmass}{m_\mathrm{p}}
\newcommand{\mppercmcube}{\protonmass \, \cmcube}
\newcommand{\hMpc}{h^{-1} \mathrm{\Mpc}}
\newcommand{\Msol}{\textup{M}_\mathrm{\sun}}

\newcommand{\xMsol}[2]{\ensuremath{{#1}\times 10^{#2} \,\Msol}}
\newcommand{\xScientific}[2]{\ensuremath{{#1} \times 10^{#2}}}

\newcommand{\kms}{\text{km} \, \text{s}^{-1}}

\newcommand{\mdm}{m_{\mathrm{DM}}}

\newcommand{\rvir}{r_{200}}
\newcommand{\Mvir}{M_{200}}

\newcommand{\etot}{\text{E}_{\text{tot}}}
\newcommand{\jz}{L_{\text{z}}}
\newcommand{\vphi}{v_{\phi}}
\newcommand{\vr}{v_{r}}

\newcommand{\zinfall}{z_{\text{infall}}}
\newcommand{\zpericentre}{z_{\text{pericentre}}}
\newcommand{\zend}{z_{\text{coalescence}}}
\newcommand{\Mstar}{M_{\star}}
\newcommand{\rhalflight}{r_{1/2, V}}
\newcommand{\feh}{\rm [Fe/H]}
\newcommand{\alphafe}{\rm [\alpha/Fe]}
\newcommand{\vrvtheta}{v_{r} / v_{\theta}}
\newcommand{\tbirth}{t_{\text{birth}}}

\title[Chemodynamical signatures of early mergers]
{VINTERGATAN-GM: The cosmological imprints of early mergers on Milky-Way-mass galaxies}

\author[M. P. Rey et al.]
{Martin P. Rey,$^{1}$\thanks{E-mail: \href{martin.rey@physics.ox.ac.uk}{martin.rey@physics.ox.ac.uk}} Oscar Agertz,$^{2}$ Tjitske K. Starkenburg,$^{3}$  Florent Renaud,$^{2}$ Gandhali D. Joshi,$^{4}$
\newauthor Andrew Pontzen,$^{4}$ Nicolas F. Martin,$^{5, 6}$ Diane K. Feuillet$^{2}$ and Justin I. Read$^{7}$
\vspace{0.8mm}
\\
$^{1}$ Sub-department of Astrophysics, University of Oxford, DWB, Keble Road, Oxford OX1 3RH, UK \\ 
$^{2}$ Lund Observatory, Department of Astronomy and Theoretical Physics, Lund University, Box 43, SE-221 00 Lund, Sweden \\
$^3$ Center for Interdisciplinary Exploration and Research in Astrophysics (CIERA), Northwestern University, 1800 Sherman Ave, Evanston, IL 60201, USA \\
$^{4}$ Department of Physics and Astronomy, University College London, London WC1E 6BT, UK \\
$^{5}$ Universit\'e de Strasbourg, CNRS, Observatoire astronomique de Strasbourg, UMR 7550, F-67000 Strasbourg, France \\
$^{6}$ Max-Planck-Institut f{\"u}r Astronomie, K{\"o}nigstuhl 17, D-69117 Heidelberg, Germany \\
$^{7}$ Department of Physics, University of Surrey, Guildford GU2 7XH, UK
}

\date{Accepted in MNRAS}

\pubyear{2022}

\begin{document}
\label{firstpage}
\pagerange{\pageref{firstpage}--\pageref{lastpage}}
\maketitle

\begin{abstract}
We present a new suite of cosmological zoom-in hydrodynamical ($\approx 20\, \pc$ spatial resolution) simulations of Milky-Way mass galaxies to study how a varying mass ratio for a Gaia-Sausage-Enceladus (GSE) progenitor impacts the $z=0$ chemodynamics of halo stars. Using the genetic modification approach, we create five cosmological histories for a Milky-Way-mass dark matter halo ($\Mvir \approx 10^{12} \, \Msol$), incrementally increasing the stellar mass ratio of a $z\approx2$ merger from 1:25 to 1:2, while fixing the galaxy's final dynamical, stellar mass and large-scale environment. We find markedly different morphologies at $z=0$ following this change in early history, with a growing merger resulting in increasingly compact and bulge-dominated galaxies. Despite this structural diversity, all galaxies show a radially biased population of inner halo stars like the Milky-Way's GSE which, surprisingly, has a similar magnitude, age, $\feh$ and $\alphafe$ distribution whether the $z\approx2$ merger is more minor or major. This arises because a smaller \textit{ex-situ} population at $z\approx2$ is compensated by a larger population formed in an earlier merger-driven starburst whose contribution to the GES can grow dynamically over time, and with both populations strongly overlapping in the $\feh-\alphafe$ plane. Our study demonstrates that multiple high-redshift histories can lead to similar $z=0$ chemodynamical features in the halo, highlighting the need for additional constraints to distinguish them, and the importance of considering the full spectrum of progenitors when interpreting $z=0$ data to reconstruct our Galaxy's past.
\end{abstract}


%%%%%%%%%%%%%%%%%%%%%%%%%%%%%%%%%%%%%%%%%%%%%%%%%%

%%%%%%%%%%%%%%%%% BODY OF PAPER %%%%%%%%%%%%%%%%%%

\section{Introduction}
The orbits and chemical abundances of stars within a galaxy encode information about its dynamical and enrichment history, providing us with a window into the main events of its cosmological formation history. The advent of the Gaia space telescope has transformed our ability to perform such analysis in the Milky Way, thanks to a dramatic improvement in the quality and volume of astrometric data sets and reconstructed stellar orbital parameters (\citealt{GAIACollaboration2016,GAIACollaboration2016DR1, GAIACollaboration2018DR2,GAIACollaboration2021eDR3, GAIACollaboration2022DR3}). Combined with chemical abundances and radial velocities acquired by large, spectroscopic surveys (e.g. \textsc{apogee}; \textsc{galah}; \textsc{h3}; \citealt{Majewski2017,Martell2017, Conroy2019}, respectively), this now allows us to isolate coherent chemodynamical structures in the solar neighbourhood and link them to ancient events several billion years back in the Milky Way's history (see \citealt{Helmi2020} for a review).

Several such clustered structures have now been identified in the high-dimensional space of orbital parameters and stellar abundances (e.g. \citealt{Belokurov2018, Helmi2018, Myeong2019, Kruijssen2020, Myeong2022}), with the most striking feature being the excess of stars on radial orbits in the local stellar halo around the Sun (the Gaia-Sausage-Enceladus; \citealt{Belokurov2018, Helmi2018}; see also \citealt{Nissen2010, Koppelman2018, Haywood2018}). This feature is most commonly interpreted as the remnant trace of our Galaxy's last significant merger, with the metallicities, ages and eccentricities of its stars pointing to a dwarf galaxy colliding with the proto-Milky Way around $z\approx2$. 

However, the exact mass-scale of the GSE progenitor remains to be pinpointed, with stellar mass estimates extending over an order of magnitude ($\Mstar \sim 10^8 - 10^9 \, \Msol$; e.g. \citealt{Bonaca2020GESages, Feuillet2020, Kruijssen2020, Mackereth2020, Naidu2020, Naidu2021GSECharacterisation, Limberg2022}). This makes it difficult to quantify its mass ratio with the proto-Milky Way and thus its impact on the early Galaxy. It also remains unclear whether the GSE feature observed at $z=0$ is a pure population that can be robustly linked to one single event, or whether it contains a superposition of multiple population with distinct origins (e.g. \citealt{Grand2020, Donlon2022Photometry, Donlon2022Chemodynamics, Myeong2022, Orkney2022GSEDouble, Khoperskov2022-1InSitu}). Furthermore, other chemokinematic debris in the disc and halo could be associated with the merger event (e.g. the `Splash'; \citealt{Bonaca2017, Haywood2018, DiMatteo2019, Belokurov2020}; `Arjuna'; e.g. \citealt{Naidu2020,Naidu2021GSECharacterisation}) but could also be of entirely distinct origin (e.g. \citealt{Amarante2020, Donlon2020, Pagnini2022}).

These uncertainties reflect the difficulties of inferring a galaxy's billions of years of dynamical and chemical evolution, from a single data snapshot. Cosmological simulations of galaxy formation provide an ideal framework for such inference, naturally providing an environment within which mergers, mass growth and subsequent star formation and chemical enrichment are self-consistently seeded from the early Universe. However, resolving the internal dynamical structure of galaxies requires large numbers of particles to adequately sample stellar phase-space orbits and avoid spurious heating ($N \gtrapprox 10^6$; e.g. \citealt{Sellwood2013,Ludlow2019DiskHeating, Ludlow2021}). Furthermore, resolving the multiphase, dense structure of the interstellar medium (ISM) from which stars form is key to accurately capture their birth kinematics and thus the subsequent dynamical structure and evolution of a disc (e.g. \citealt{House2011, Bird2013}). 

Recent progress in numerical methods and computing power now allow modern cosmological zoom simulations to meet these requirements within individual Milky-Way-mass galaxies, enabling us to model a handful of objects sampling varying environments and formation scenarios (e.g. \citealt{Sawala2016, Grand2017, Grand2021, Buck2020NIHAOUHD, Font2020, Applebaum2021, Agertz2020Vintergatan, Bird2021, Khoperskov2022-1InSitu, Wetzel2022}). With such tools, we can now quantify the frequency of chemodynamical patterns at $z=0$, link them to specific events in each galaxy's history, and inform the reconstruction of our Galaxy's formation. (e.g. \citealt{Bignone2019, Mackereth2019, Fattahi2019, Elias2020, Dillamore2022, Khoperskov2022-2Accreted, Khoperskov2022-3Metallicity, Pagnini2022}).

However, a causal interpretation of such suites of individual galaxies still remains challenging. The formation scenario of each given galaxy originates from stochastic early-Universe perturbations, making two simulated galaxies' merger histories entirely unrelated to one another. It then becomes difficult to assess how specific $z=0$ chemodynamical signatures would respond to a change in the early cosmological merger history, and whether such signatures are unique to this formation scenario or can be produced through multiple routes. This fundamentally limits the robustness with which we can reconstruct our Milky Way's past, and our understanding of the degeneracies associated with this inference.

In this work, we address this challenge by combining two approaches: (i) high-resolution ($\approx 20 \, \pc$), cosmological zoomed simulations using a physical model that can successfully reproduce the chemodynamical structure of Milky-Way-like galaxies (\citealt{Agertz2020Vintergatan, Renaud2020Vintergatan2, Renaud2020Vintergatan3}); and (ii) the genetic modification approach (\citealt{Roth2016, Rey2018, Stopyra2021}). Genetic modifications allow us to create different versions of a chosen cosmological galaxy, introducing targeted changes to its formation scenario such as the mass ratio of a specific merger (e.g. \citealt{Pontzen2017, Sanchez2021}), the overall formation time (e.g. \citealt{Rey2019UFDScatter, Davies2021MorpholicalTrans}) or the angular momentum accretion (e.g. \citealt{Cadiou2022}). Each modified version differs minimally from the original scenario, for example conserving the total dynamical mass at $z=0$ and the large-scale environment around the galaxy. This enables controlled, comparative studies in a fully cosmological context, isolating how a specific aspect of the formation scenario affects the final observables of a galaxy.     

Specifically, in this work, we target a Milky-Way-inspired merger history, studying a $\approx 10^{12} \, \Msol$ dark matter halo which experiences a merger at $z\approx2$ on a radially biased orbit similar to that inferred for a potential GSE progenitor. Using the genetic modification approach, we then make the merger mass ratio incrementally smaller and larger to create a suite of five related formation scenarios, all with similar cosmological environment and all converging to the same total dynamical mass to within 10 per cent (\citealt{Rey2022}). Each of these cosmological scenarios is then evolved to $z=0$ using simulations that resolve the galaxy's ISM multiphase structure and include the detailed star formation, stellar feedback and chemical enrichment model used for the \textsc{vintergatan} project (\citealt{Agertz2020Vintergatan}). The five genetically modified galaxies used in this work further form a subset of a larger suite evolved with this model, which will be described in a forthcoming work (Agertz et al. in preperation).

We present the simulation suite in Section~\ref{sec:methods}, and show in Section~\ref{sec:morphologies} that modifying the mass ratio of the GSE-like event at $z\approx2$ leads to markedly distinct galactic morphologies at $z=0$, at fixed dynamical and stellar mass. Despite this structural diversity and large variations in mass ratios, we obtain similar GSE-like phase-space features at $z=0$, whose stars have similar median ages and $\feh$ distributions (Section~\ref{sec:chemodynamics}). We discuss the consequences of our findings on inferring merger mass ratios from Galactic data in Section~\ref{sec:discussion} and summarize in Section~\ref{sec:conclusion}.

\section{Genetically-modified Milky-Way galaxies} \label{sec:methods}
We present and analyse a suite of five genetically modified, Milky-Way-mass galaxies, systematically varying the significance of an early $z\approx2$ merger similar to the inferred properties of the GSE progenitor. A thorough description of how we construct genetically modified, cosmological zoomed initial conditions to vary the significance of this event is available in \citet{Rey2022}, while the physical model used to evolve them to $z=0$ is described in \citet{Agertz2020Vintergatan}. We summarize these aspects in Section~\ref{sec:sec:simulations}, and describe the controlled changes to the merger scenarios of each galaxy introduced by our modifications in Section~\ref{sec:sec:mergerscenarios}.

\subsection{Numerical setup and galaxy formation physics} \label{sec:sec:simulations} 

We construct cosmological, zoomed initial conditions using the \textsc{genetic} software (\citealt{Stopyra2021}) and cosmological parameters $\Omega_{m} = 0.3139$, $h = 0.6727$, $\sigma_8 = 0.8440$, $n_s = 0.9645$ (\citealt{PlanckCollaboration2016}). Starting from a dark matter-only cosmological volume with a box size $50  \, \hMpc \approx 73 \, \Mpc$ and mass resolution $\mdm = \xMsol{1.2}{8}$, we select a target reference halo with Milky-Way virial mass ($\Mvir \approx 10^{12} \, \Msol$) and no neighbours more massive within $5\, \rvir$, where $\rvir$ is the radius enclosing 200 times the critical density of the Universe. We then refine the mass resolution in the Lagrangian region down to $\mdm =  \xMsol{2.0}{5}$, and apply the procedure described in \citet{Pontzen2021} to strongly damp the bulk velocity of the Lagrangian region and limit advection errors during integration. The initial conditions are evolved using linear theory to $z=99$ (\citealt{Zeldovich1970}), before we start following the evolution of dark matter, stars, and gas to $z=0$ with the adaptive mesh refinement code \textsc{ramses} (\citealt{Teyssier2002}). 

We follow the dynamics of collisionless particles (dark matter and stars) using a multiscale particle-mesh solver (\citealt{Guillet2011}), while fluid dynamics are solved with an HLLC Riemann solver (\citealt{Toro1994}) assuming an ideal gas equation of state with adiabatic index $\gamma = 5/3$. Our Lagrangian refinement strategy allows us to reach a spatial resolution of $20 \, \pc$ throughout the galaxy's ISM (\citealt{Agertz2020Vintergatan}). We complement this hydrodynamical setup with an extensive galaxy formation model described in detail in \citet{Agertz2020Vintergatan}, which we briefly summarize now.

We follow the equilibrium cooling of a metal-enriched plasma (\citealt{Courty2004, Rosdahl2013}), and model the spatially uniform, time-dependent heating and photoionization from a cosmic ultraviolet background using an updated version of \citet{Haardt1996} as implemented in the public \textsc{ramses} version. Gas with $n_{\textsc{H}} \geq 0.01 \, \cmcube$ self-shields from this heating source (\citealt{Aubert2010, Rosdahl2012}), allowing it to condense to densities $\rho \geq 100 \, \mppercmcube$ and temperatures $T \leq 100$ K at which we model star formation through a Poisson process following a Schmidt law (\citealt{Schmidt1959, Rasera2006, Agertz2013}). Newborn stellar particles are sampled with $10^4 \, \Msol$ masses and are treated as single stellar populations with a \citet{Chabrier2003} initial mass function.

We track the age-dependent injection of momentum, energy, and metals from stellar winds in O, B, and asymptotic giant branch (AGB) stars, and explosions of Type II and Type Ia supernovae (SNe) according to the budget defined in \citet{Agertz2020Vintergatan} (see also \citealt{Agertz2013,Agertz2020EDGE, Agertz2015}). Feedback from SNe is injected as thermal energy if the cooling radius is resolved by at least 6 grid cells, and as momentum otherwise (\citealt{Kim2015, Martizzi2015, Agertz2020EDGE, Agertz2020Vintergatan}). We track the evolution of two metals, iron and oxygen, using progenitor-mass--dependent stellar yields for SNe (\citealt{Woosley2007}), a delay-time distribution for SNeIa (`field' in  \citealt{Maoz2012}), and a continuous slow, release from AGB stars (\citealt{Agertz2015}).

We identify dark matter haloes and subhaloes using the \textsc{AHF} structure finder (\citealt{Gill2004,Knollmann2009}), retaining only structures with more than 100 particles. We construct merger trees by matching haloes and subhaloes across each simulation snapshot using the \textsc{pynbody} (\citealt{Pontzen2013}) and \textsc{tangos} (\citealt{Pontzen2018}) libraries. Halo centres are identified using the shrinking-sphere algorithm (\citealt{Power2003}), and we shift the coordinate frame to ensure that velocities within the central kpc vanish. We define the total stellar mass of each galaxy, $\Mstar$, by summing the stellar mass within $\rvir$, and we interpolate a single stellar population model (\citealt{Girardi2010}) over a grid of ages and metallicities to obtain the luminosities of individual stellar particles. The projected half-light radii, $\rhalflight$ is then derived along a random line of sight. We define the iron and $\alpha$ abundance ratios following \citet{Agertz2020Vintergatan}, equation (3) (see also \citealt{Escala2018}) converting to solar ratios using \citet{Asplund2009}, table 1, and using oxygen as an approximation for the total $\alpha$ abundance as it dominates the mass fraction. We compute total metallicities from these two elements following \citet{Kim2014}, equation (4).

\begin{figure}
  \centering
    \includegraphics[width=\columnwidth]{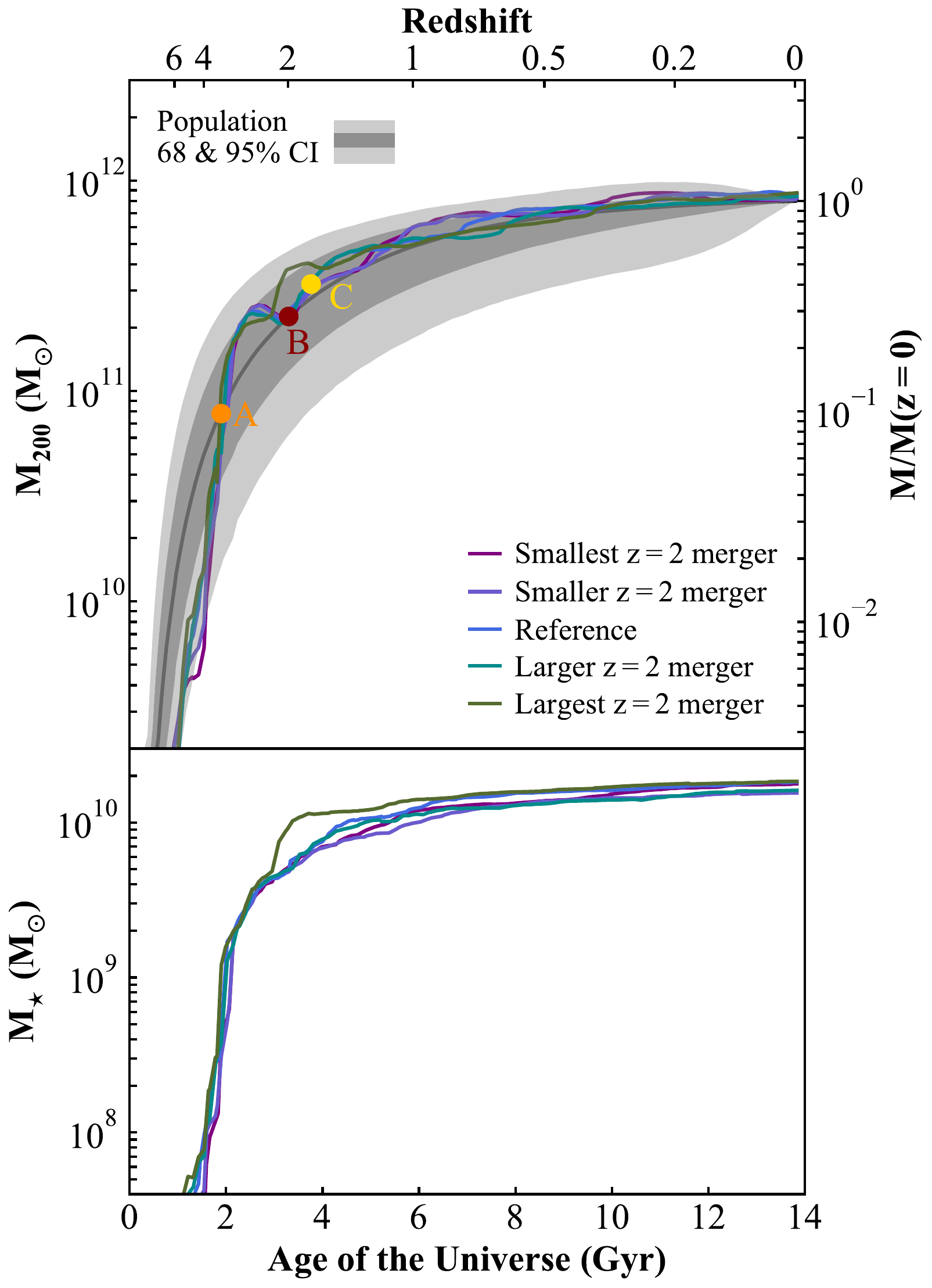}

    \caption{Dynamical (top) and stellar (bottom) mass assemblies over cosmic time across the suite of genetically modified galaxies. The significance of a merger at $z\approx2$ (B, red dot) in the reference case (blue line) is incrementally decreased (navy and purple lines) and increased (cyan and green) using genetic modifications. By construction, the final dynamical mass of all galaxies is fixed, inducing small compensating shifts in an early major event (A, orange) and a late minor merger (C, yellow, see Table~\ref{table:mergers} for all merger properties). All galaxies converge to similar stellar masses at $z=0$ (bottom) and evolve in similar large-scale environment (Figure~\ref{fig:redshift3}).
    }
    \label{fig:massgrowths}
\end{figure}

\subsection{Genetic modifications and merger scenarios} \label{sec:sec:mergerscenarios} 

\begin{figure*}
  \centering
  \includegraphics[width=\textwidth]{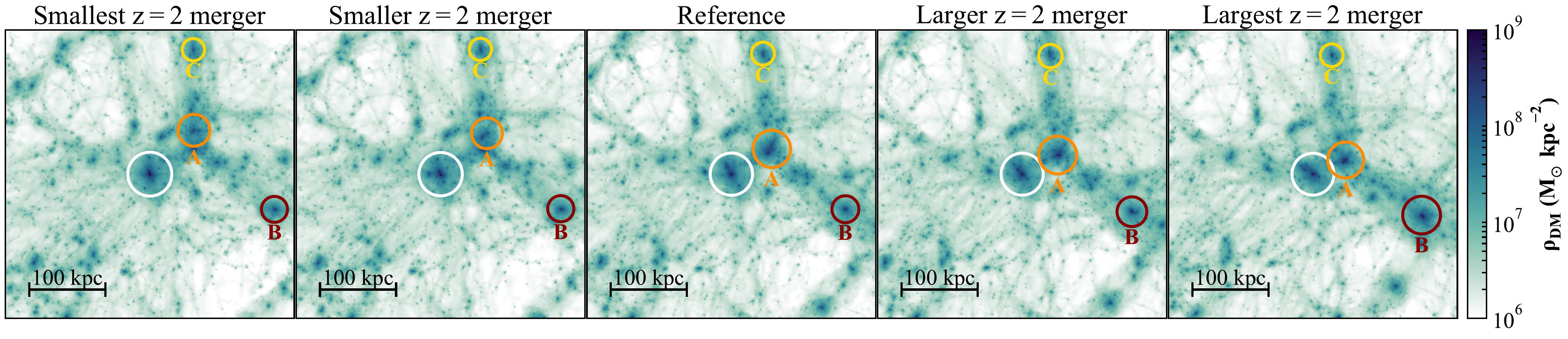}
  \caption{Projected dark matter density maps at $z=3.5$ showing the large-scale environment of each merger scenario. As we enhance the significance of the $z\approx2$ merger (`B', red) using genetic modifications, its progenitor becomes visibly larger (left to right, circles show $\rvir$), while the cosmological filamentary geometry is maintained in each case. Due to correlations and compensations to conserve the total mass, other events (`A' and `C', orange and yellow, respectively) see their infall time and merger ratios slightly modified (see Table~\ref{table:mergers} for a complete quantification).}

  \label{fig:redshift3}
\end{figure*}

\begin{table*}
  \centering
  \caption{Properties of the three main merger events (A, B, and C, marked in Figures~\ref{fig:massgrowths} and~\ref{fig:redshift3}) cross-matched across each genetically modified scenario (individual columns). For each event, we quote the redshift at which the infaller's centre is last outside the main body's $\rvir$, $\zinfall$. At this time, we report the progenitor's virial mass, stellar mass, mass ratios with the central body and radial to tangential velocity ratio at $\rvir$. We identify the redshift of the interaction's first pericentre passage and final coalescence (see Section~\ref{sec:sec:mergerscenarios}). Self-consistently cross-matching structures between simulations can lead to mass ratios less than unity if the identity of the major progenitor swaps at very early times (merger `A' in first two columns).} 

    \begin{tabular}{c c c c c c}
    \hline
    & Smallest $z=2$ merger & Smaller $z=2$ merger & Reference & Larger $z=2$ merger & Largest $z=2$ merger \\
    \hline

    \color{BrickRed}{Target progenitor B} & $\zinfall = 1.94$ & $\zinfall = 2.03$ & $\zinfall = 1.99$ & $\zinfall = 1.99$ & $\zinfall = 2.24$ \\

    Infall $\Mvir$ & $\xMsol{2.2}{10}$ & $\xMsol{2.3}{10}$& $\xMsol{3.8}{10}$& $\xMsol{7.4}{10}$& $\xMsol{1.1}{11}$ \\

    Infall $\Mstar$ & $\xMsol{2.2}{8}$ & $\xMsol{3.1}{8}$& $\xMsol{6.0}{8}$& $\xMsol{1.2}{9}$& $\xMsol{2.3}{9}$ \\

    Ratio $\Mvir$ & 1:10 & 1:9.8  & 1:6.0 &  1:2.9 &  1:2.1 \\
    Ratio $\Mstar$ & 1:24 & 1:15  & 1:8.1 &  1:4.3 &  1:2.1 \\

    Infall $\vrvtheta$ & 9.5 & 9.1 & 16. &  8.8 &  5.5 \\

    $\zpericentre$ & $1.70$ & $1.76$ & $1.79$ & $1.80$ & $1.98$ \\

    $\zend$ & $1.40$ & $1.45$ & $1.52$ & $1.47$ &  $1.65$ \\

    \hline

    \color{orange}{Earlier progenitor A} & $\zinfall=3.11$ & $\zinfall=3.17$ & $\zinfall=3.35$ & $\zinfall=3.35$ & $\zinfall=3.49$ \\

    Infall $\Mvir$ & $\xMsol{9.6}{10}$ & $\xMsol{7.7}{10}$& $\xMsol{5.8}{10}$& $\xMsol{5.1}{10}$& $\xMsol{3.7}{10}$ \\

    Infall $\Mstar$ &$\xMsol{6.2}{8}$ & $\xMsol{5.2}{8}$& $\xMsol{4.5}{8}$& $\xMsol{3.8}{8}$& $\xMsol{3.1}{8}$ \\

    Ratio $\Mvir$ & 1:0.7 & 1:0.9  & 1:1.3 &  1:1.4 & 1:1.9 \\

    Ratio $\Mstar$ & 1:1.8 & 1:2.1  & 1:2.4 & 1:1.8 & 1:2.4 \\

    Infall $\vrvtheta$ & 9.5 & 4.1 & 2.3 &  2.3 &  2.0 \\

    $\zpericentre$ & $2.74$ & $2.72$ & $2.86$ & $2.94$ & $3.00$ \\

    $\zend$ &  $2.17$ & $2.30$ & $2.34$ & $2.34$ &  $2.38$ \\

    \hline

    \color{Goldenrod}{Later progenitor C} & $\zinfall = 1.94$ & $\zinfall = 1.86$ & $\zinfall = 1.73$ & $\zinfall = 1.70$ & $\zinfall = 1.62$ \\

    Infall $\Mvir$ & $\xMsol{1.5}{10}$ & $\xMsol{1.8}{10}$& $\xMsol{1.8}{10}$& $\xMsol{1.8}{10}$& $\xMsol{1.7}{10}$ \\

    Infall $\Mstar$ & $\xMsol{1.7}{8}$ & $\xMsol{2.8}{8}$& $\xMsol{2.3}{8}$& $\xMsol{2.8}{8}$& $\xMsol{2.2}{8}$ \\

    Ratio $\Mvir$ & 1:15 & 1:14  & 1:18 &  1:19 &  1:22 \\
    Ratio $\Mstar$ & 1:31 & 1:20 & 1:29 &  1:26 &  1:50 \\

    Infall $\vrvtheta$ & 0.5 & 0.6 & 0.5 &  0.4 &  0.6 \\

    $\zpericentre$ & $1.61$ & $1.59$ & $1.48$ & $1.41$ & $1.35$ \\
    $\zend$ & $0.09$ & $0.04$ & $ 0.17$ & $0.22$ & $0.20$ \\
    \hline
    \end{tabular}
   \label{table:mergers}
\end{table*}

The genetic modification technique makes targeted changes to a galaxy's cosmological initial conditions to modify its later non-linear merger history in a controlled way. Modifications and initial conditions used in this work are extensively described in \citet{Rey2022} in a dark-matter-only context (see their `Milky-Way-like' family) - here, we re-evolve this same family of initial conditions with hydrodynamical galaxy-formation simulations (Section~\ref{sec:sec:simulations}) and now present their resulting merger scenarios.

Our study targets a dark matter halo with a total dynamical mass $\Mvir \approx \xMsol{1}{12}$ and which experiences an early, large interaction with a radially biased orbit (infall at $z\approx2$ with a $\Mvir$ ratio of 1:6 and a radial-to-tangential velocity ratio between the progenitors $\vrvtheta = 16$ at this time; see `Target Progenitor' in Table~\ref{table:mergers}). We choose this history to broadly resemble the inferred properties of the progenitors of the proto-Milky Way (\citealt{Belokurov2018, Helmi2018}). Using genetic modifications, we then aim to increase and decrease the significance of this event, bracketing the range of reported progenitor masses and mass ratios. We thus identify the Lagrangian patch of this early merger in the reference object and define linear modifications to increase or decrease its mean overdensity and control the merger mass ratio (see \citealt{Rey2022}, table 1 for the detailed modifications). 

Figure~\ref{fig:massgrowths} shows the assembly of dynamical (top) and stellar (bottom) mass in the main progenitor of each merger scenario. Our modifications successfully conserve the early ($z\geq 5$) and late ($z\leq 0.5$) accretion histories, converging by design to similar dynamical masses at $z=0$ to within 10 per cent (\citealt{Rey2022}). Further, Figure~\ref{fig:redshift3} shows that the large-scale environment around the galaxy is largely unchanged between scenarios. This is a natural consequence of the genetic modification approach which aims to make minimal changes to variables untargeted by the modifications while retaining consistency with the $\Lambda$CDM cosmology (see also \citealt{Pontzen2017, Rey2019VarianceDMOs} for further visuals). 

Mass growth histories around $z\approx2$ however diverge, as expected following our explicit targeting of a merger event at this time. Before quantifying these changes, we briefly assess the likelihood of each genetically modified accretion history to verify any potential rarity in a $\Lambda$CDM universe. The relative likelihood of each new modified initial condition to the reference, $\Delta \chi^2$, remain small compared to the available number of degrees of freedom -- $\Delta \chi^2 = -3.6$, $-2.77$, $-0.45$, $+0.3$ for increasing merger mass ratio compared to $\approx10^6$ modes in the zoom region -- ensuring their compatibility with the $\Lambda$CDM power spectrum. We further compute the median and 68-95\% fractional mass growth histories across a population of 28,475 dark matter haloes extracted from the IllustrisTNG simulation (\citealt{Nelson2019}, grey contours in Figure~\ref{fig:massgrowths}, see \citealt{Rey2022} for further details on the computation). All our mass growth histories are within the 1$\sigma$ contour of the overall population around $z\approx2$, demonstrating that the scenarios studied in this work are all plausible cosmological realizations (see also Section~\ref{sec:discussion:likelihood} for further discussion on compatibility with $\Lambda$CDM merger rates). 

To quantify differences in merger histories, we extract merger events in the reference scenario that (i) bring at least $\Mstar \geq 10^8 \, \Msol$ of accreted material at infall and with (ii) merger mass ratios more significant than 1:30 in $\Mvir$. These cuts ensure that we eliminate both very high-redshift events with high mass ratios but low significance to the overall content, and late, low-redshift mergers with small mass ratios. This flags three key progenitors in the reference merger history (labelled `A', `B'. and `C' in Figures~\ref{fig:massgrowths} and~\ref{fig:redshift3}), which, as we will see in Section~\ref{sec:chemodynamics} all play a role in defining the $z=0$ chemokinematic structure within our galaxies. 

We cross-match these merger events across our genetically modified simulations, and report in Table~\ref{table:mergers} their infall redshifts, defined at the time at which the infaller's centre is last outside the main progenitor's virial sphere, and their masses and mass ratios at this time. We further compute the ratio between the radial and tangential velocities of the two halo centres at this time to assess the angular momentum of the encounter, and identify the times of first pericentric passages, $\zpericentre$, and end of the interaction, $\zend$, using the high-cadence simulation movies.

Our genetic modifications explicitly target merger `B' (second row in Table~\ref{table:mergers}; red in Figures~\ref{fig:massgrowths} and~\ref{fig:redshift3}), which in the reference case is the last major event until $z=0$. We make it incrementally more significant in each scenario in both mass ratio, dynamical and stellar mass (Table~\ref{table:mergers}), while retaining a strongly radial approach in all cases ($\vrvtheta \geq 5$). The range of merging stellar masses scanned by our suite ($\xMsol{2}{8} \leq \Mstar \leq \xMsol{2}{9}$) accurately brackets the proposed masses for a potential GSE progenitor (e.g. \citealt{Belokurov2018, Helmi2018, Feuillet2020, Kruijssen2020, Mackereth2020, Naidu2020, Naidu2021GSECharacterisation, Limberg2022}).

Performing these targeted changes to merger `B' however leads to other alterations to the merger history, due to the correlations inherent to a cosmological context (\citealt{Roth2016, Rey2018}). For example, reducing the mass of an event (e.g. `B') while maintaining the same total mass at $z=0$ is compensated by growing other events (here mainly `A', Table~\ref{table:mergers}). Further genetic modifications could force the mass ratio of the earlier event `A' to match across all scenarios and control for this effect. We leave such finer control to future work, and stress that having multiple significant, high-redshift progenitors is a generic prediction of $\Lambda$CDM (see Section~\ref{sec:discussion:likelihood}) and is thus inherently reflected by the cosmological nature of the genetic modification approach. 

Another effect of our modifications is to slightly alter the infall times of each merger as their significance and mass varies (Table~\ref{table:mergers}). This is most visible in Figure~\ref{fig:redshift3}, where merger `A' (orange) is already overlapping with the main progenitor (white) in the rightmost panel, but is further away in the leftmost case. Such shifts in infall time arise from correlations between the linear density, velocity, and potential fields in the initial conditions - as we increase or decrease the local overdensity to modify the merger mass ratio, it smooths or sharpens the local potential gradient towards the main progenitor, in turn modifying the velocity field and introducing shifts in merger timings (see \citealt{Pontzen2017, Rey2019VarianceDMOs} for further examples). Again, infall times could be fixed to their reference values using additional modifications targeting the velocity or angular momentum structure of the Lagrangian patch (e.g. \citealt{Cadiou2021AngMomGM, Pontzen2021}). However, the differences in timings ($\approx 300 \, \Myr$) remain small compared to uncertainties in dating such early merger events in our Milky Way (e.g. \citealt{Bonaca2020GESages, Feuillet2021}), and we thus leave a more detailed setup simultaneously controlling infall times and merger ratios to a future study. 

By design of our modifications, all host dark matter haloes converge to the same dynamical mass at $z=0$, but we note that their central galaxies also match in their final stellar mass ($\Mstar = \xMsol{1.8}{10}$ within 20 per cent of each other; Figure~\ref{fig:massgrowths}). All final stellar masses are compatible at 1$\sigma$ with empirical model predictions for this halo mass (\citealt{Moster2018, Behroozi2019}), and we will show in a forthcoming work that their growth over time is also compatible with such models. We interpret this convergence in stellar mass as a byproduct of the extremely similar dynamical mass assemblies and large-scale environments of each genetically modified galaxy. 

To summarize, our five genetically modified scenarios provide us with a controlled study that systematically varies the significance of an early, radially infalling merger at $z\approx2$, while fixing the large-scale cosmological environment, and the total dynamical and stellar mass budget of a Milky-Way-mass galaxy.

\section{Response of the galaxies' structure to growing an early merger} \label{sec:morphologies}

\begin{figure*}
  \centering
  \includegraphics[width=\textwidth]{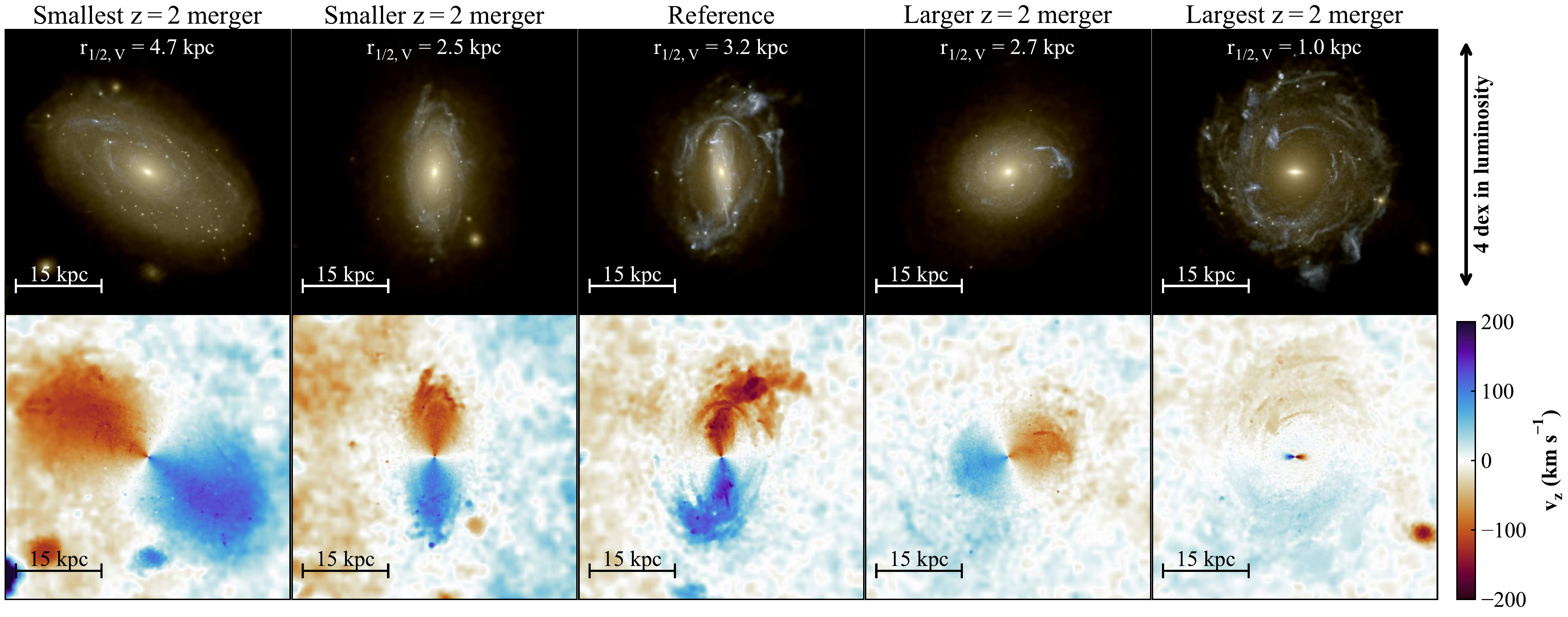}

  \caption{Response of the $z=0$ stellar light (top) and kinematics (bottom) to modifying our galaxies' early histories. As we grow the $z\approx2$ merger (left to right), the galaxy transitions from an extended, rotationally supported disc (left-most), to a more compact, bulge-dominated structure (right-most). All changes occur at fixed galaxy stellar and dynamical mass (Figure~\ref{fig:massgrowths}) and in similar large-scale environment (Figure~\ref{fig:redshift3}), showcasing the diversity in galaxy morphology and kinematics introduced by different cosmological histories. Despite this diversity, all scenarios forming a well-ordered galactic disc (first four columns) showcase a GSE-like phase-space structure (Figure~\ref{fig:phasespace}, \ref{fig:vrvphi}), that has distinct cosmological origin but overlapping chemodynamics and ages in every case (Figure~\ref{fig:radialstarsSFR}, \ref{fig:radialstarsFeH}).
  }

  \label{fig:overview}
\end{figure*}

We start by quantifying the response of the $z=0$ stellar structure of each galaxy as we genetically modify their early assembly history. Figure~\ref{fig:overview} shows UVI-mock images of the stellar light at $z=0$ (top row) and luminosity-weighted line-of-sight velocity maps (bottom row) as we incrementally increase the significance of the $z\approx2$ merger (left to right). Images do not account for dust attenuation, span a surface brightness interval from 20 to $30 \, \text{mag arcsec}^{-2}$ and are viewed along the simulation's z-axis (i.e. a random line of sight physically, but a consistent orientation across panels).

Starting from the reference case (central panels), the galaxy showcases an irregular morphology at $z=0$, with a rotationally supported, inner galactic disc which extends into a bluer corotating outer structure misaligned with the inner stars. As we make the early history of the galaxy quieter (from central to left-hand panels), the disc orders into a single kinematic component, and spatially grows. By contrast, a more significant merger increases the compactness of the galaxy, forming increasingly brighter central bulges and reducing the overall rotational support (fourth and fifth panels). 

The galaxy with the largest merger (right-most panels) exhibits an inner core, counterrotating compared to an outer, low surface brightness ($\geq 27 \, \text{mag arcsec}^{-2}$) star-forming disc. Such structure is reminiscent of the decoupled internal kinematics observed around local elliptical galaxies (e.g. \citealt{Efstathiou1982, Bender1988, Krajnovic2011, Krajnovic2015, Johnston2018, Prichard2019}), for which such low-surface brightness discs have been revealed by deep imaging (e.g. \citealt{Duc2015}). This galaxy stands out in our suite compared to the more ordered, rotationally supported objects (first four columns), and is difficult to relate to studies of our Galaxy. We present it here as a useful complement to establish trends with growing mass ratio, and will provide a dedicated study of its internal kinematics in a follow-up work.

Our results highlight a broad trend: quieter histories favour larger stellar discs (left), while early major mergers produce more compact, bulge-dominated morphologies at $z=0$ (right and Section~\ref{sec:chemodynamics}). This trend aligns with expectations that mergers drive, in part, morphological transformations of galaxies (e.g. \citealt{Negroponte1983, DiMatteo2007, Hopkins2009, Naab2014, Zolotov2015, Pontzen2017, Martin2018, Davies2021MorpholicalTrans}). We cleanly isolate this effect here, as our transformations in galactic structure occur at fixed dynamical masses, fixed galaxy stellar masses, and similar large-scale cosmological environments (Figures~\ref{fig:massgrowths} and ~\ref{fig:redshift3}).

However, our different histories at fixed mass showcase further diversity in galactic morphological and kinematic structure than this broad trend. The effective radii (labelled in each panel) are modified by a factor five from left to right, and do not respond linearly to the growth of the early $z\approx2$ merger (e.g. second, third, and fourth columns decrease and increase in turn). We verified that this is also the case for 3D stellar half-mass radii, which are less sensitive to mass-to-light assumptions. Moreover, each galaxy's disc significantly varies in orientation with respect to the same frame of reference in Figure~\ref{fig:overview}.   

This is to be expected as other physical variables that affect disc formation are evolving across our genetically modified scenarios. In particular, the spin-orbit coupling, the impact parameter, and the resulting tidal fields in the $z\approx2$ interaction vary between scenarios and can play a key role in setting the bulk of the angular momentum of the future galaxy (e.g. \citealt{Hernquist1993, Springel2005DiscFormation, Hopkins2009, Renaud2009, Renaud2020Vintergatan2}). Furthermore, later aspects of the galaxy's history ($z\leq1$) relevant to disc formation are also modified, notably the constructiveness of filamentary gas accretion (e.g. \citealt{Dekel2020, Kretschmer2020}), the structure of the inner CGM and its angular momentum content (e.g. \citealt{Stern2021, Hafen2022, Gurvich2023}), or the orbital parameters and spin of late minor mergers (e.g. \citealt{Jackson2022MinorMergers}). 

None the less, despite this structural diversity in the inner galaxy, we will see in Section~\ref{sec:chemodynamics} that the chemodynamics of halo stars in each galaxy remain largely similar. We hypothesize that the misaligned discs in the reference scenario highlight a `turning point' in cosmological angular momentum accretion, with these two discs getting constructively aligned when diminishing the early merger to produce larger discs, and destructively counteraligned when growing the early merger, leading to more compact morphologies (e.g. \citealt{Kretschmer2020}). We leave a confirmation of this scenario to a future study, and now focus on extracting the signatures of each merger scenario in phase-space.

\section{Chemodynamical signatures of high-redshift merger histories} \label{sec:chemodynamics}

\subsection{Phase-space structure and radially biased halo stars} \label{sec:chemodynamics:ges}

\begin{figure*}
  \centering
    \includegraphics[width=\textwidth]{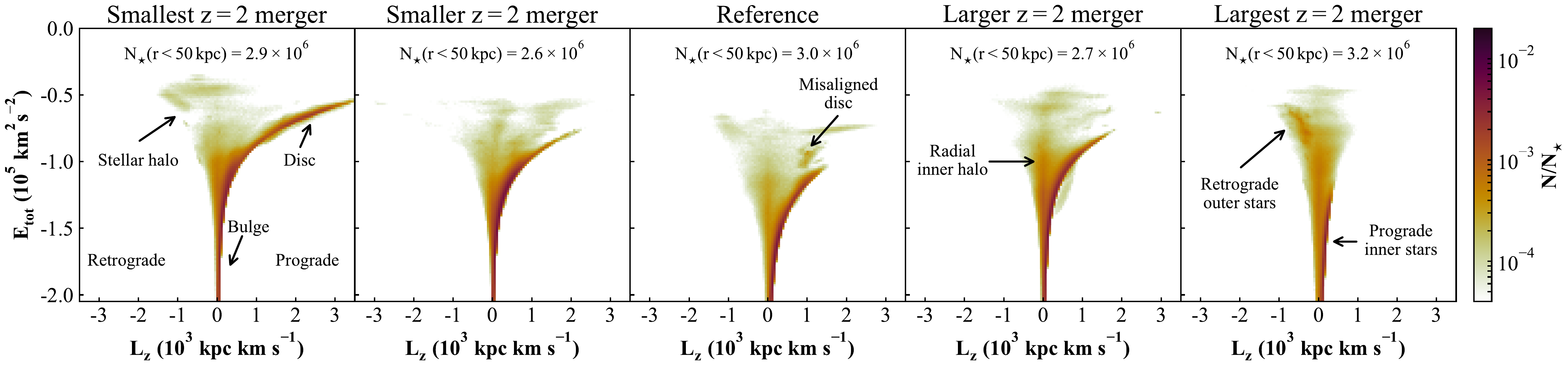}

    \caption{Fractional distributions of specific energy and vertical angular momentum for all stars within $50$ kpc at $z=0$, as we increase the significance of the early merger (left to right). Broad trends and specific galactic structures identified in Figure~\ref{fig:overview} are recovered in phase space (annotated), in particular the increasingly smaller stellar discs with increasing merger ratio (e.g. lack of a high-$\jz$ sequence in the right-most panel). A population of inner halo stars (low $\jz$, $\etot$ comparable to disc stars) is present in all merger scenarios, but becomes more prominent compared to the high-$\jz$ disc population with increasing merger mass ratio (fourth panel). This population of inner halo stars is strongly radially biased, resembling the GSE population in the Milky Way (Figure~\ref{fig:vrvphi}). Clustered structures in the outer halo (e.g. left-hand panel) map to dwarf galaxy satellites and recently disrupted debris.
    }
    \label{fig:phasespace}
\end{figure*}

\begin{figure*}
  \centering
    \includegraphics[width=\textwidth]{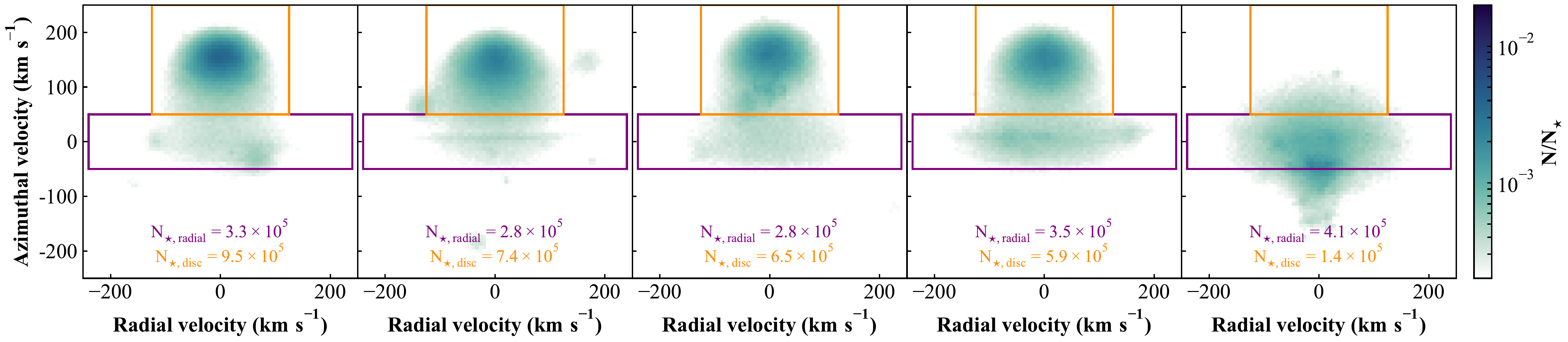}

    \caption{Azimuthal and radial velocity distribution of stars with $2 \leq r \leq 50$ kpc, as we increase the significance of the early merger (left to right). For well-ordered disc galaxies (first four columns), the velocity distribution is bimodal, with disc stars (positive $\vphi$, orange boxes) linking into a radially biased population ($\vr / \vphi \gg 1$, purple box) alike the Milky-Way's GSE structure. We detect this latter population in all cases, in similar absolute numbers (purple), despite the diversity in final galactic structure and past merger mass ratio at $z\approx2$. Further to these kinematic similarities, radially biased GSE-like stars exhibit similar age and $\feh$ distributions in all cases (Figures~\ref{fig:radialstarsSFR} and~\ref{fig:radialstarsFeH}).
    }
    \label{fig:vrvphi}
\end{figure*}

To construct the chemodynamical structure of each galaxy, we start by extracting the phase-space distribution of all stars within the inner $50 \, \kpc$. Figure~\ref{fig:phasespace} shows the fractional distributions of their specific vertical angular momentum, $\jz$, and orbital energies, $\etot$. We define the vertical disc direction from the parallel to the stellar angular momentum in the inner 5 kpc, and compute $\etot$ by summing the kinetic energy of each stellar particle with the local gravitational potential computed by the simulations' Poisson solver. This accounts for asphericity, time-dependence, and long-range cosmological fluctuations in the potential - we thus re-normalize it to have vanishing $\etot$ at the virial radius and ease comparison with isolated dynamical studies. The total number of stars normalizing each histogram is shown at the top of each panel of Figure~\ref{fig:phasespace}, and changes by less than 20 per cent across scenarios. This follows the minimal changes between each galaxy's total stellar mass (Section~\ref{sec:sec:mergerscenarios}) and ensures that comparisons between fractional phase-space distributions are minimally affected by different normalizations.

Starting from the `smallest $z=2$ merger' scenario (left-hand panel), we can first identify different galactic components in phase-space: (i) the bulge, with bound orbits (low $\etot$) and pressure-supported kinematics (low $\jz$); (ii) the stellar disc, as a sequence spreading across a range of $\etot$ and extending towards rotationally supported, high-$\jz$ orbits; and (iii) the fainter stellar halo surrounding the galaxy, towards higher $\etot$ and small $\jz$. The clustered, coherent structures in the stellar halo's phase space map onto disrupted debris on long dynamical time-scales and surviving dwarf galaxy satellites around each galaxy.

Each of these generic components can be mapped across the different formation scenarios, recovering the visual and kinematic structural trend highlighted in Figure~\ref{fig:overview}. The disc sequence in phase space is the most extended in $\jz$ for the quietest merger history (left-hand panel), significantly smaller for intermediate scenarios (second, third, and fourth columns) and lacking for the largest merger (right-hand panel). Furthermore, we can identify in phase-space the specifics kinematic structures noted previously. The misaligned stellar discs in the reference scenario (central panel) appear as a double sequence, with an inner, $\jz$-tail linking to the bulge, and a population overlapping in $\jz$ but at higher $\etot$. Similarly, the decoupled kinematics for the largest merger (right-hand panel) are visible as a minimal prograde disc sequence for the inner stars (small $\etot$), and retrograde, higher $\etot$ orbits for the outer stars. 

Beyond these already noted features, the phase-space distributions exhibit a significant population of inner halo stars, with pressure-supported orbits (low $\jz$) that share the same range of $\etot$ as disc stars. This population is visible in all galaxies but becomes particularly prominent for the larger merger mass ratios (e.g. fourth column), as expected if a single early event dominates the assembly of the inner halo. In fact, Figure~\ref{fig:phasespace} exhibits a systematic trend: increasing the mass ratio of an early merger makes the inner halo population more prominent relative to disc stars (darker colours compared to the disc stars from left to right).

To quantify this trend in the stellar halo in more detail, we extract the kinematics of the stellar population, excluding the innermost, bulge-dominated stars (i.e. stars with $2 \leq r \leq 50\, \kpc$) and show their azimuthal and radial velocity fractional distributions in Figure~\ref{fig:vrvphi}. All merger scenarios with a well-ordered disc component (first four columns; the right-most galaxy showcases counterrotating inner and outer kinematics, hence its misalignement in this plot) exhibit a bimodal kinematic distribution, with (i) a population of stars with positive azimuthal velocities rotating with the disc (highlighted by the orange boxes), which links into (ii) a population of stars with small azimuthal velocities over a large radial extent (purple boxes). This latter population of radially biased stars is highly reminiscent of the GSE population in the Milky Way (e.g. \citealt{Belokurov2018, Helmi2018}), which is key to claims that our Galaxy underwent a large merger around $z\approx2$. 

\begin{figure*}
  \centering
    \includegraphics[width=\textwidth]{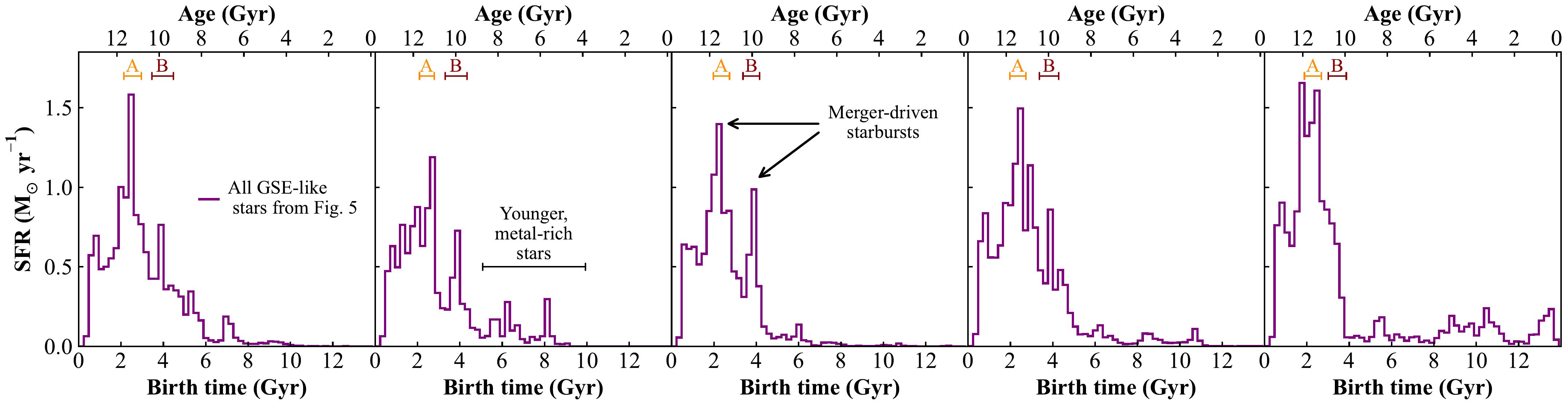}

    \caption{Star formation histories of GSE-like stars identified in Figure~\ref{fig:vrvphi} for each merger scenario. Their ages show extended distributions, with similar medians ($ \approx 11\, \Gyr$) across all scenarios consistent with an ancient population. Interactions with progenitors `A' and `B' (interaction timings marked at the top) map onto isolated bursts of formation of GSE-like stars, indicating that a significant fraction of this population originates from early gas-rich mergers. The lack of such correspondence for merger `B' and its largest merger mass ratio (right-most panel) suggests however that the origin of stars on GSE-like orbits is evolving across merger scenarios (Figure~\ref{fig:radialstarsFeH}). A younger, metal-rich, $\alphafe$-reduced population contaminates the sample of ancient populations due to our broad kinematic cuts to define GSE-like stars (see Section~\ref{sec:chemodynamics:originradialdebris:metals} and Appendix~\ref{app:extradata} for further characterization).
    }
    \label{fig:radialstarsSFR}
\end{figure*}

However, we recover this orbital structure in each of our galaxies with ordered rotation (first fourth columns), whether our merger at $z\approx2$ is enhanced to become a major event (stellar mass ratio up to 1:4; fourth panel) or decreased to a minor event (down to a merger ratio less than 1:20; left-most). We further extract the number of stars on radial orbits by integrating the distribution within the purple boxes and quote it in each panel. Despite their diversity in early merger history and final galactic structure, all four galaxies have nearly identical absolute number of stars on radial orbits, changing by at most thirty per cent across scenarios. This is comparable to changes in overall stellar mass (numbers in Figure~\ref{fig:phasespace}) and to variations when observing the galaxy at different nearby timestamps (Section~\ref{sec:chemodynamics:radialdebrisevolution}).
 
We can further quantify the relative contribution of inner halo stars compared to disc stars defined within the orange boxes in Figure~\ref{fig:vrvphi} (orange number). The fraction of radial-to-disc stars follows a systematic trend with $z\approx2$ merger mass ratio (left to right), going from 0.34 to 0.37 to 0.43 to 0.59 for the four disc galaxies, and to 2.9 in the last scenario which lacks well-ordered rotation. We checked that this comparative trend holds when selecting stars with $r \geq 1 \, \kpc$ and $r\geq 3 \, \kpc$, or focusing only on halo-like orbits with $| \jz | \leq 0.5 \, \kpc\, \kms$. 

Our results thus demonstrate that GSE-like inner halo populations can be assembled through both minor and major events at $z\approx2$, and might be relatively common across the Milky-Way analogue population (see Section~\ref{sec:discussion:likelihood} for further discussion). Further, the overall mass of a GSE-like population does not directly link to the mass ratio of a single, early merger. Rather, a larger merger systematically increases the relative contrast between this population and the galactic disc in phase-space, at fixed galaxy stellar mass and dynamical mass. We thus caution that inferring the merger mass ratio and mass-scale of the GSE progenitor requires quantifying relative weights between subpopulations (e.g. disc to inner halo) which, in turn, calls for a thorough understanding of the completeness and selection functions of both the observed and simulated data (see Section~\ref{sec:discussion:mw} for further discussion). This weak connection between GSE-like structures at $z=0$ and the mass ratio of a $z\approx2$ merger questions the cosmological origin of these stars, and we now rewind each progenitor's history to establish their exact nature.

\subsection{The origin of radially biased halo stars} \label{sec:chemodynamics:originradialdebris}

\subsubsection{Ages and metallicity distributions} \label{sec:chemodynamics:originradialdebris:ages}

We now focus on testing the cosmological origin of the GSE-like stars identified kinematically in Figure~\ref{fig:vrvphi} (purple box with $| \vphi | \leq 50 \, \kms$, $| \vr | \leq 240 \, \kms$, and $2 \leq r \leq 50$ kpc). We start by plotting their star formation histories in Figure~\ref{fig:radialstarsSFR} (merger significance increasing from left to right). This does not account for stellar mass-loss across cosmic time, but we checked that our results are unchanged if plotting the mass-weighted distribution of stellar ages.

In all cases, ages show a broad distribution, with most stars being older than 9 Gyr ($z\geq 1.7$), as expected if they track ancient stellar populations associated to early merger events. However, despite the order-of-magnitude variation in the merger mass ratio at $z\approx2$, all GSE-like populations have similar median ages -- 11.2, 11.3, 11.5, 11.3, 11.4 Gyr for each panel, respectively -- which would be hard to distinguish from data once convolved with realistic uncertainties ($\approx 1 \, \Gyr$; e.g. \citealt{Feuillet2016, Bonaca2020GESages, Xiang2022}).

A tail of later forming stars is also visible in all cases ($\tbirth \geq 6 \, \Gyr$), reflecting a contamination from younger stars with overlapping kinematics at $z=0$. We will confirm in Section~\ref{sec:chemodynamics:originradialdebris:metals} and Appendix~\ref{app:extradata} that these are young, \feh-rich, $\alphafe$-poor stars consistent with being formed within the disc and in satellite dwarf galaxies that infall later into the main progenitor. Such contamination could likely be minimized through optimized chemical and dynamical selections tailored to each galaxy, but we rather choose broader cuts to provide consistent and complete comparisons between scenarios.

\begin{figure*}
  \centering
    \includegraphics[width=\textwidth]{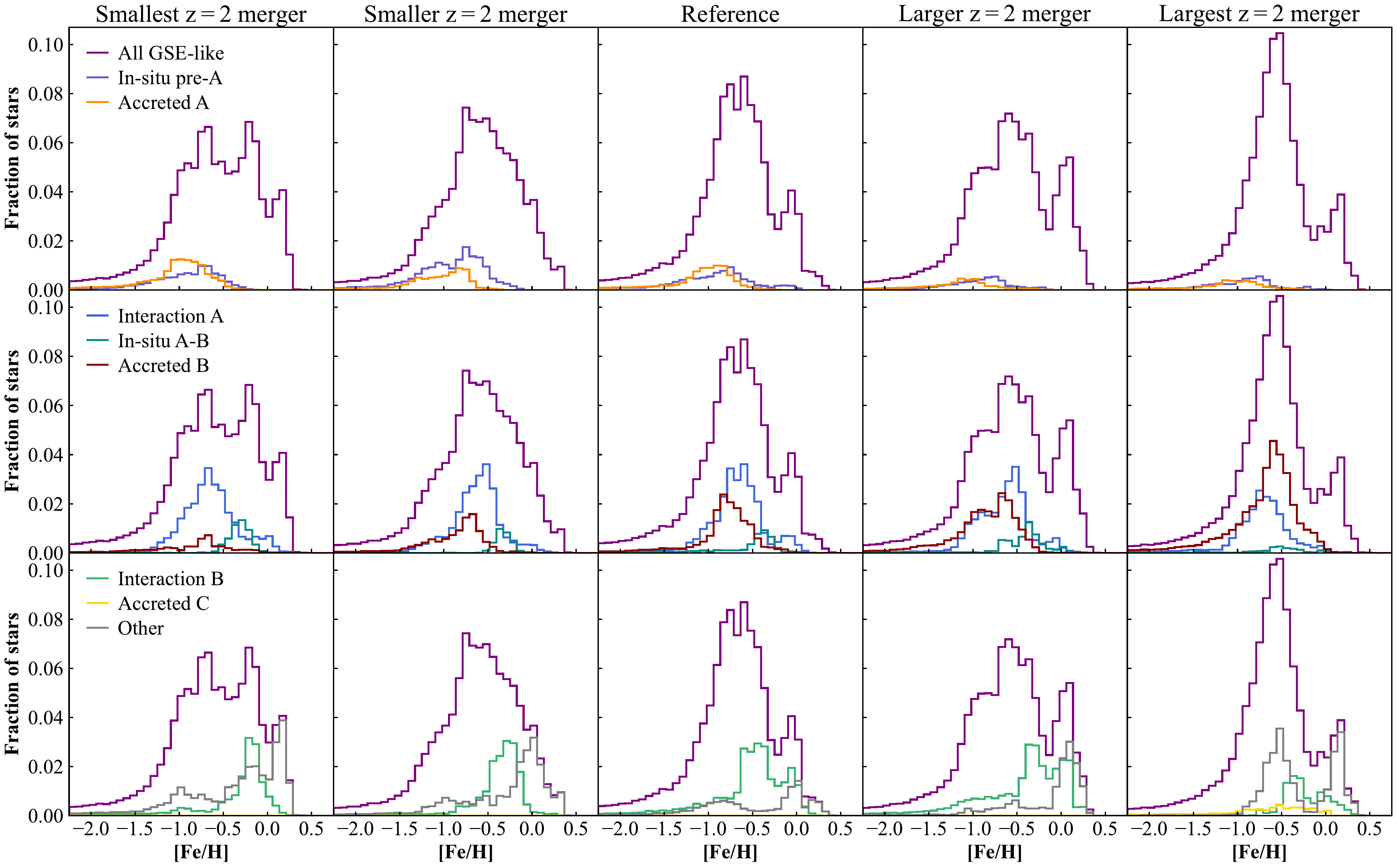}

    \caption{Iron metallicity distribution of the GSE-like stars in each merger scenario (left to right), broken down into subpopulations linked to early merger events (coloured lines, formed before interaction `A', between interactions `A' and `B' and afterwards, in top, middle and bottom rows respectively). In all cases, the overall population (purple) shows extended $\feh$ distributions, with similar medians despite the wide-scan in mass ratio for the GSE-like merger progenitor. The origin of the stars however shifts between merger scenarios. The accreted component of the GSE-like progenitor (`B', red, second row) grows with its mass ratio (left to right), but stars formed during the interaction with an earlier merger (`A'; light blue, second row) always significantly contribute and have overlapping $\feh$. A small mass ratio at $z\approx2$ (left-hand panels) is thus compensated by the growth of this earlier-formed population, resulting in similar chemodynamics and magnitude for the GSE feature at $z=0$ at fixed galaxy stellar and dynamical mass. Earlier populations (top row) contribute little to the overall population but are significant for the metal-poor tail of the distribution. Later populations (bottom row) represent a significant fraction of the total, but could be easily filtered due to their higher metallicities and lower $\alphafe$ (Figure~\ref{fig:debris2Dchemodynamics}).}
    \label{fig:radialstarsFeH}
\end{figure*}

Furthermore, we show in Figure~\ref{fig:radialstarsFeH} the $\feh$ distributions of the GSE-like stars (purple). Again, all scenarios exhibit broad distributions of metallicities reminiscent of that observed in the Milky Way's (e.g. \citealt{Feuillet2020, Naidu2021GSECharacterisation}), that all peak around a similar median -- $\langle\feh \rangle= -0.61$, $-0.61$, $-0.70$, $-0.59$, $-0.58$ from left to right panels, respectively. We thus conclude that the order-of-magnitude scan in merger mass ratio at $z\approx2$ not only produces kinematic similarities (Figure~\ref{fig:vrvphi}), but also similar distributions of ages and metallicities for GSE-like stars. 

To understand the nature of these similarities between merger scenarios, we note that the star formation histories of the radially biased halo population at $z=0$ also exhibit singled-out enhancements compared to the background distribution. We mark at the top of Figure~\ref{fig:radialstarsSFR} the length of the interaction with mergers `A' and `B' (see Table~\ref{table:mergers} for $\zinfall$ and $\zend$\footnote{We omit merger `C' for clarity, as, despite its early infall, its long inspiral which lasts until $z=0$ does not drive a starburst.}). For all regular disc galaxies (first four panels), peaks of star formation in the GSE-like population coincide with interactions, and we show in Appendix~\ref{app:starburst} that such peaks further map to galaxy-wide starbursts in the overall stellar population. Such interaction-driven starbursts have been recently detected in the star formation history of the Milky Way, where peaks of star formation coincide with probable pericentre passages of the Sagittarius dwarf galaxy (e.g. \citealt{Ruiz-Lara2020}). Starbursts triggered by early, gas-rich mergers thus play a significant role in assembling the GSE-like population at $z=0$, indicating that a significant fraction of its stars are of \textit{in-situ} origin (see also \citealt{Grand2020, Orkney2022GSEDouble, Khoperskov2022-1InSitu}) and could elevate the galaxy-wide star formation rate $\approx10 \, \Gyr$ ago (e.g. \citealt{Alzate2021MWSFH}).

Furthermore, the long tails towards low ($\feh \leq -1.0$) and high ($\feh \geq -0.5$) metallicities, and the spread in ages, imply that multiple populations of different origins are populating the $z=0$ GSE-like population. In fact, despite having the most massive merger at $z\approx2$, the last scenario (right-most) lacks a correspondence between the interaction of `B' and a peak of star formation. We show in Appendix~\ref{app:starburst} that this merger does in fact trigger a large, galaxy-wide, star formation enhancement, but that stars formed during this burst do not populate GSE-like orbits at $z=0$. The nature of radially biased halo stars and the respective importance of each interaction is thus evolving across the different merger scenarios,  which we quantify now.

\subsubsection{Multiple co-exisiting in-situ and ex-situ populations} \label{sec:chemodynamics:originradialdebris:metals}

To establish the respective origin of stars and the role of each merger event in assembling the $z=0$ GSE-like feature, we track each GSE-like star through the cosmological merger trees and assign them to eight mutually exclusive categories:
\begin{enumerate}[align=left, font=\bfseries]
  \item[In-situ pre-A:] Stars that are within the main progenitor before the first infall of merger `A' (Table~\ref{table:mergers}). We define a star's membership to a dark matter halo from the halo finder, and verified that defining it from spheres of $\rvir$ and $2 \, \rhalflight$ around the halo centre does not affect the relative trends reported in this work.
  \item[Accreted A:] Stars that are within `A' at its first infall, which form a clean sample of \textit{ex-situ} stars from this event.
  \item[Interaction A:] Stars that are within the main progenitor between the first infall of merger `A' and its $\zend$. Stars in this subpopulation primarily form during the starburst driven by this interaction (recall the peak in star-formation in Figure~\ref{fig:radialstarsSFR}) and can originate from gas within the main host, from gas within `A' or from the gas tail produced by the interaction. We avoid distinguishing between \textit{in-situ} and \textit{ex-situ} origins for this population, as this distinction becomes ever more difficult as the two merging bodies coalescence (although see e.g. \citealt{Cooper2015}).
  \item[In-situ A-B:] Stars that are within the main progenitor between the end of merger `A' and the first infall of `B'. These are primarily \textit{in-situ} stars, although a small, sub-dominant population of \textit{ex-situ} stars deposited by very minor mergers between `A' and `B' also contributes. 
  \item[Accreted B:] Stars that are within `B' at its first infall.
  \item[Interaction B:] Stars within the main progenitor between the first infall of merger `B' and its $\zend$. 
  \item[Accreted C:] Stars within `C' at its first infall.
  \item[Other] Stars which have not been assigned to the previous categories. As we will see, these are primarily young, $\feh$-rich, low-$\alphafe$ stars formed in the disc, or deposited later in the halo from other less significant mergers than `A', `B' and `C' (see also Appendix~\ref{app:extradata}). 
\end{enumerate}

Figure~\ref{fig:radialstarsFeH} shows the respective contribution of each subpopulation to the total $\feh$ distribution of GSE-like stars (top). We further show each subpopulation's respective distributions in $\etot$ and $\alphafe$ in Appendix~\ref{app:extradata}. 

In all merger scenarios, a varying mixture of the subpopulations identified above contributes to the metallicity distributions, each peaking at distinct $\feh$ but always with broad and overlapping spread. The `Interaction A' and `Interaction B' (blue and green, middle and bottom row, respectively) subpopulations are particularly prominent, confirming that merger-driven starbursts produce a significant fraction of the total $z=0$ GSE-like population. 

\begin{figure*}
  \centering
    \includegraphics[width=\textwidth]{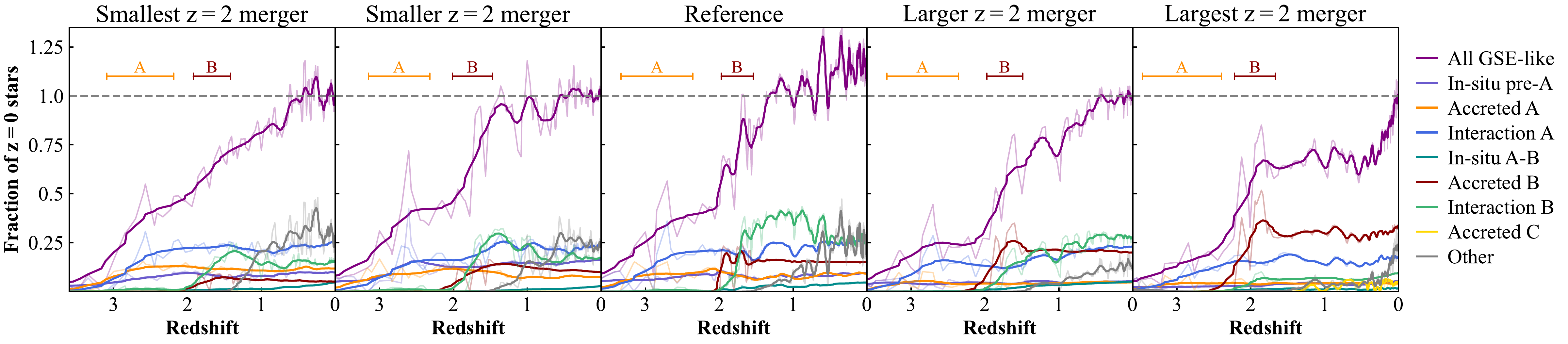}
    \caption{Build-up over time of each subpopulation within the GSE-like feature (colours matching Figure~\ref{fig:radialstarsFeH}, thick lines are 3-pixel Gaussian smoothings of full-data thin lines). The timing of high-redshift interactions (`A' and `B' shown at the top) mark the build-up of their associated populations. The composition in ancient populations of the GSE-like feature observed at $z=0$ is already largely in place at $z=1$ (all colours but the younger, contaminating grey population). Larger mass ratios of merger `B' (centre and right-hand panels) show tentative `Splash'-like population, visible as an increased number of \textit{in-situ} stars on radial orbits following the interaction (blue), but this remains small compare to time-step-to-time-step noise and later dynamical evolution.}
    \label{fig:debrisbuildup}
\end{figure*}

As we increase the importance of merger `B' (left to right), its accreted population (red, middle row) rises. It only dominates, however, in the most extreme scenario (right-most), for which the galaxy transitions into a bulge-dominated morphology (Figure~\ref{fig:overview}). In fact, when the final galaxy showcases organized rotation at $z=0$ (first four columns), the population of stars formed during the earlier interaction `A' (blue, middle row) dominates the overall budget of radial stars for small mass ratios (1:24 and 1:15; first two columns), and has nearly perfectly overlapping $\feh$ distribution with the accreted population (red) for medium mass ratios (1:8 and 1:4; third and fourth columns). Our results thus reaffirm that multiple components associated to several early accretion events can mix and contribute to a galaxy-wide, metal-poor radial population (see also \citealt{Grand2020, Donlon2022Photometry, Donlon2022Chemodynamics, Orkney2022GSEDouble}).

Further, within each individual merger scenario, distinguishing such subpopulations chemically could be highly ambiguous. One might expect a population of stars formed in the central regions during the interaction to be more metal-rich than the accreted component associated to the merger body. We recover this trend for each galaxy, where the accreted component of a given interaction peaks toward lower $\feh$ (orange versus blue; red versus green). However, telling apart an accreted component at $z\approx2$ from an earlier \textit{in-situ} population is difficult, as they peak around the same metallicity and exhibit extended distributions that overlap with one another (blue versus red; see also \citealt{Renaud2020Vintergatan2}). We further show in Appendix~\ref{app:extradata} that, similarly to $\feh$, they also overlap in $\etot$ and $\alphafe$ and their 2D combinations.

Furthermore, these ambiguities strongly propagate to potential inferences of the merger mass ratio at $z\approx2$. For the four disc galaxies (first four columns), diminishing the mass ratio at fixed total stellar and dynamical mass (recall Figure~\ref{fig:massgrowths}) reduces the accreted contribution from this merger, but is compensated by growing the earlier \textit{in-situ} population. This systematic variation in origin of the $z=0$ GSE-like stars still results in similar overall $\feh$ distributions, combining with their overlapping star formation histories (Figure~\ref{fig:radialstarsSFR}), kinematics (Figure~\ref{fig:vrvphi}), $\alphafe$, and $\etot$ (Appendix~\ref{app:extradata}). 

Our results thus highlight strong degeneracies when inferring the early merger history of a galaxy from its chemodynamics, with multiple histories leaving overlapping signatures at $z=0$. We stress that there likely remains powerful avenues to distinguish each merger scenario, in particular leveraging additional chemical information and precise stellar ages (e.g. \citealt{Myeong2019, Bonaca2020GESages, Feuillet2021,Matsuno2022SequioaAbundances}). None the less, we show that a robust inference of our Galaxy's past require a wider scan through merger histories than previously explored, which accounts for the full spectrum of high-redshift progenitors and the response of the protogalaxy to each of them, combined with a careful account of their observables in the solar neighbourhood. We discuss this latter aspect further in Section~\ref{sec:discussion:mw} and focus next on quantifying when the GSE-like structure and its subpopulations are assembled across cosmic time.

\subsection{Are stars born, deposited, or kicked onto radial orbits?} \label{sec:chemodynamics:radialdebrisevolution} 

Our results establish that \textit{in-situ} populations forming in earlier interactions overlap kinematically and chemically with later accreted populations in the inner $z=0$ halo. However, it remains unclear whether all subpopulations are born on halo-like radial orbits and retain memory of their formation. Or whether they acquire their radial orbits later during the galaxies' history, for example due to subsequent interactions or secular evolution.

To answer these questions, we repeat the analysis performed in Sections~\ref{sec:chemodynamics:ges} and~\ref{sec:chemodynamics:originradialdebris} along the cosmological merger tree of each galaxy. Briefly, we extract face-on azimuthal and radial velocities at each saved snapshot and apply the same cuts as in Section~\ref{sec:chemodynamics:ges} to isolate the GSE-like population ($| \vphi | \leq 50 \, \kms$, $| \vr | \leq 240 \, \kms$, and $2 \leq r \leq 50$ comoving kpc). We then repeat the assignment performed in Section~\ref{sec:chemodynamics:originradialdebris} to identify subpopulations and show their respective numbers over time in Figure~\ref{fig:debrisbuildup}. We verified that all trends discussed in this section are qualitatively unchanged when plotting the number of stars which are both in a population at a given time and at $z=0$. We further observe significant variation from one time-step to the next due to shifts in galaxy centres and orientation affecting our kinematic and spatial cuts. Figure~\ref{fig:debrisbuildup} thus shows the full time evolution data and a 3-pixel Gaussian-smoothed curve for readability (thin and thick lines, respectively). 

We show on Figure~\ref{fig:debrisbuildup} the timings of the `A' and `B' interactions ($\zinfall$ to $\zend$ marked at the top). As expected, these time periods coincide with the rise of the specific subpopulations associated to these events. For all merger scenarios, the respective contributions of ancient populations is roughly in place at $z\approx1$, after the most significant interactions have completed. The contaminating, younger-forming population grows later into the GSE-like feature (grey), consistent with contamination from disc stars and later-accreted populations (Section~\ref{sec:chemodynamics:originradialdebris} and Appendix~\ref{app:extradata}).

Interaction populations (blue and green) exhibit the strongest time evolution, although it appears as a continuous process rather than being linked to localized triggers. In particular, the passage of `B' does not correlate with a significant increase in older populations (violet, orange, blue, and green) on GSE-like orbits, as would be expected if the merger dynamically kicks a significant population of existing central stars onto radial orbits. Such increase is tentatively visible for the larger merger mass-ratios (centre and two right-most panels), but the time-step-to-time-step variability makes a robust interpretation challenging and changes around this time are comparable to later dynamical evolution. We further verified that the galaxy always has time to dynamically relax between `A' and `B', despite what appears as two mergers in quick succession (e.g. right-most panel) -- in all scenarios, when `A' has coalesced, the dynamical time $\sqrt{2 \, r^{3} / G \, M(<r)}$ at $2 \, \rhalflight$ is shorter than 20 Myr, whereas the first infall of `B' happens at the fastest 150 Myr later. 

We thus conclude that merger `B' plays a key role in (i) delivering accreted stars onto radial orbits (red), (ii) forming stars on radial orbits by driving a starburst during its interaction (green), (iii) but minimally modifies the orbits of existing GSE-like stars. 

Rather, dynamical evolution appears more gradually, as a slow increase in the number of stars on GSE-like orbits over time, which is particularly apparent for \textit{in-situ} populations within our disc galaxies (blue and green lines in first four panels). These components can see their numbers double between $z=1$ and $z=0$ and overtake the more stable, accreted components. Our results thus stress the importance of capturing the full dynamical history of a galaxy when interpreting its $z=0$ observables, as late evolution can significantly affect the composition and mixture of GSE-like features.

This slow evolution of the structure put in place at high-redshift points towards secular dynamical mechanisms within the galaxy, or repeated small perturbations over a Hubble time. We expect a general heating of old-stars orbits over time, either driven by physical perturbations such as satellite flybys (\citealt{Sellwood2002, Quillen2009}) and scattering off gas and stellar clumps (e.g. \citealt{vanDonkelaar2022, Wu2022DiscFromScattering}), or driven by numerical limitations due to the finite particle numbers and force resolution (e.g. \citealt{Ludlow2019DiskHeating}). However, it remains unclear whether such mechanisms would produce preferentially radially biased, halo-like orbits, motivating future cosmological studies quantifying their respective importance.

\section{Discussion} \label{sec:discussion}

\subsection{Disentangling high-redshift merger scenarios from a mock solar neighbourhood} \label{sec:discussion:mw}

Our results in Section~\ref{sec:chemodynamics} focus on galaxy-wide chemodynamical patterns to provide a complete and consistent comparison between merger scenarios. However, the data sets motivating our analysis, and with which we ultimately wish to compare, are only available within our Galaxy and around the particular location of the Solar neighbourhood. Furthermore, these data sets have non-trivial spatial and kinematic selection functions, which could potentially bias inferences of the merger history compared to galaxy-wide trends. Transforming the stellar populations of cosmological simulations into resolved stars to account for such selection functions requires careful, dedicated modelling (e.g. \citealt{Grand2018Aurigaia, Sanderson2020, Thomas2021, Lim2022}), and is outside the scope of this work. In this section, we none the less provide a preliminary assessment of whether the multiple radially-biased halo populations highlighted in Section~\ref{sec:chemodynamics} would remain as overlapped chemodynamically in a mock Galactic survey.

Our simulated galaxies are less massive than the Milky Way at $z=0$ ($\Mstar \approx \xMsol{2}{10}$) and lack some of its defining late-time features such as a bar (Figure~\ref{fig:overview}) and a Local Group environment. This limits direct, absolute comparisons with Galactic data, but still allows us to establish comparative trends between scenarios. We thus focus on our two intermediate scenarios `Smaller $z=2$ merger' and `Larger $z=2$ merger' that are closest to our Galaxy: they both show a regular disc structure necessary to define an equivalent Solar neighbourhood, have optical sizes compatible with the Milky Way's thin disc ($\approx 2.5 \, \kpc$, e.g. \citealt{McMillan2017}).

To mimic Galactic surveys of halo stars, we then centre on an arbitrary point in the disc plane at 8 kpc from the galactic centre and select stars with inclinations $|b| > 15^{\circ}$ and distances $D < 20 \, \kpc$. We then repeat the same kinematic cuts as in Section~\ref{sec:chemodynamics:ges} to isolate the GSE-like population and perform the same assignment procedure as in Section~\ref{sec:chemodynamics:originradialdebris} to link stars with early merger events. 

\begin{figure}
  \centering
    \includegraphics[width=\columnwidth]{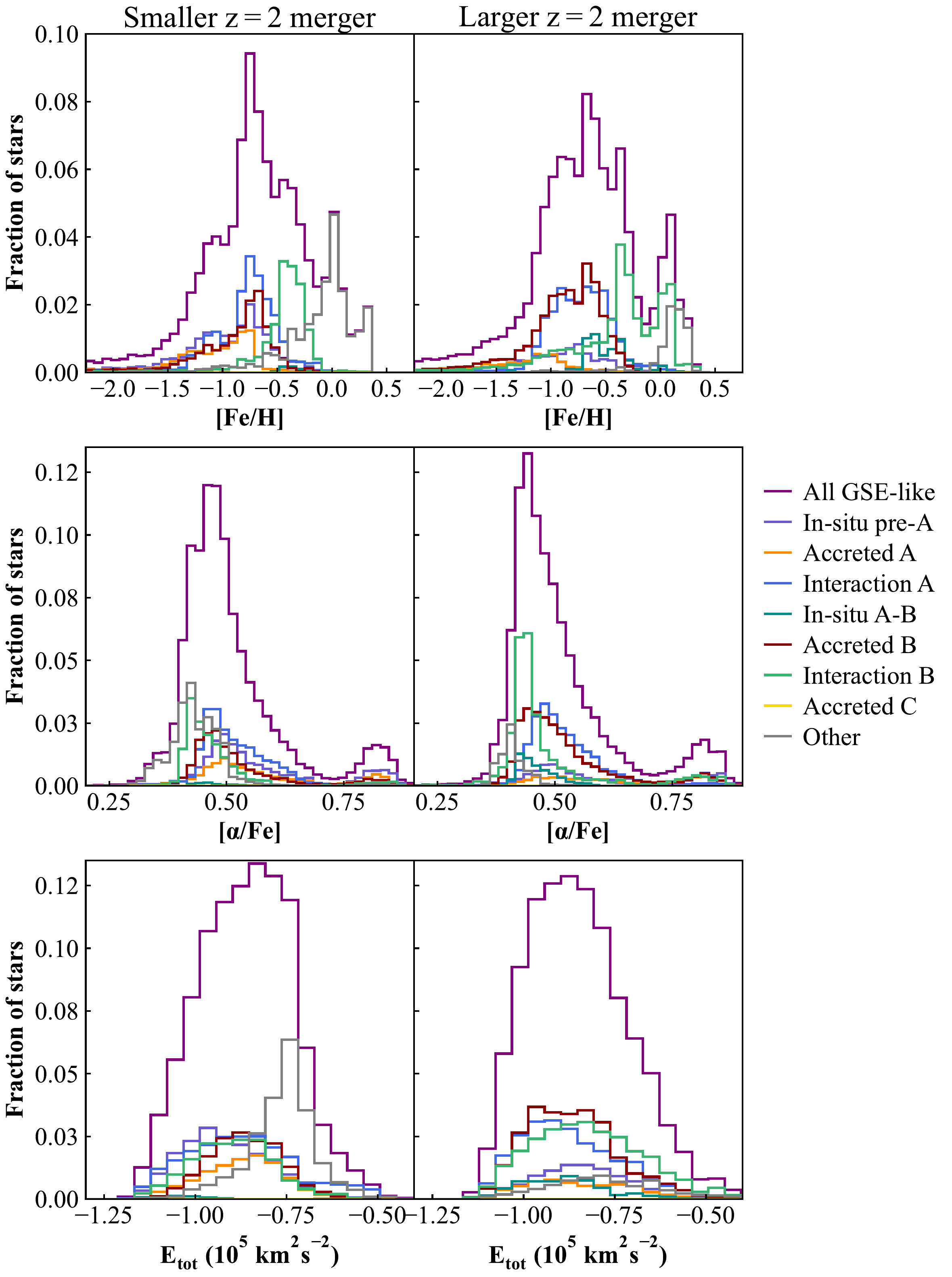}
    \caption{$\feh$, $\alphafe$, and $\etot$ distributions (top to bottom) of kinematically identified GSE-like stars in a mock Galactic halo survey from the Solar neighbourhood. With this selection function, as with galaxy-wide samples, the magnitude of the GSE-like feature is conserved for both a small and large merger mass ratio at $z\approx2$ (1:10 and 1:4 in left- and right-hand columns, respectively) and overall chemodynamical distributions strongly overlap (purple). In both cases, the overall population is a mixture of several subpopulations (colours match Figure~\ref{fig:radialstarsFeH}), which do not separate clearly in chemical or phase-space to unequivocally identify each scenario.}
    \label{fig:mwselectionfunction}
\end{figure}

Figure~\ref{fig:mwselectionfunction} shows the resulting distributions of the total GSE-like population (purple) and each of its subcomponent (coloured lines matching Figure~\ref{fig:radialstarsSFR}) in $\feh$, $\alphafe$\footnote{Using the same stellar evolution model as in \citet{Agertz2020Vintergatan}, we also recover higher $\alphafe$ ratios than observed in the Milky Way (see their Section 4 for further discussion).}, and $\etot$ (top to bottom). With this tailored selection function, we recover the same trends and conclusions as with galaxy-wide samples: (i) despite a factor-three difference in their $z\approx2$ merger's stellar and dynamical mass, both scenarios have similar total number of GSE-like stars (38,664 and 41,828 for left and right respectively); (ii) the radial population of halo stars, at both low and mid metallicities ($\feh \leq -1.0$ and $ -1.0 < \feh \leq -0.5$) is a mixture of multiple \textit{in-situ} and \textit{ex-situ} components originating from several high-redshift interactions; (iii) these populations do not separate clearly in $\alphafe$ and $\etot$, posing a challenge to the reconstruction of early merger properties from $z=0$ data.

A promising avenue for future work to distinguish these early merger scenarios could be to focus on the low-$\feh$ tail of halo stars. The absolute number of stars with $\feh \leq -1.0$ ($\feh \leq -0.5$) in the mock solar neighbourhood does not vary significantly between the two scenarios, with 9677 stars and 10343 for the left- and right-hand columns (25,024 and 27,783 stars, respectively). But the different nature and timing of the progenitors could be potentially distinguished using specific peaks in the metallicity distribution functions, combined with additional chemical abundances sensitive to different enrichment time-scales (e.g. Al and Na; \citealt{Belokurov2022Aurora, Feuillet2022}, although see e.g. \citealt{Buck2021} for uncertainties related to chemical enrichment). We plan in future work to post-process our simulations to produce such additional abundances and study in more details the patterns of metal-poor stars across our merger scenarios (see also \citealt{Sestito2021}). For now, we caution that linking $z=0$ metallicities in a GSE-like feature to the mass ratio of a single GSE-progenitor event is a delicate task. Robust inferences require broader exploration of cosmological merger histories to understand their potential degeneracies, and a careful account of their observables' in the Solar Neighbourhood to disentangle and weight the respective importance of mixed subpopulations.

\subsection{How common are two high-redshift merger events for Milky-Way-mass dark matter haloes?} \label{sec:discussion:likelihood}

Our results highlight that multiple merger events at early times play a key role in shaping the final population of radially biased, halo stars at $z=0$. In particular, Section~\ref{sec:chemodynamics:originradialdebris} highlights the important contribution of stars formed during an earlier interaction at $z\approx3$ (merger `A'), which is difficult to separate from \textit{ex-situ} stars brought with the radial event at $z\approx2$ explicitly targeted by our genetic modifications (merger `B'). In Section~\ref{sec:sec:mergerscenarios} and Figure~\ref{fig:massgrowths}, we show that the mass growths of major progenitors across all merger scenarios are within the 2$\sigma$ population from $z\approx6$. However, mass assembly can be achieved through both mergers and smooth accretion, and we now wish to assess the likelihood of having two significant mergers at such early times in a $\Lambda$CDM universe. 

To answer this question formally, one should estimate the joint likelihood of two merger events of different mass ratios one after the other, across the population of Milky-Way-mass dark matter halo in a $\Lambda$CDM universe. However, such likelihood is not readily available due to the large parameter space that needs to be sampled across all possible $\Lambda$CDM merger histories. Instead, we extract the summary statistics introduced by \citet{Fakhouri2010} from the Millenium simulation suite (\citealt{Springel2005Millenium, Boylan-Kolchin2009}) to describe the average population merger rate of dark matter haloes:
\begin{equation}
  \frac{\dd N}{\dd \xi \dd z}(\Mvir, \xi, z) = A \, (\frac{\Mvir}{10^{12} \, \Msol})^{\alpha} \,  \xi^{\beta} \, e^{ (\frac{\xi}{\xi_{0}})^{\gamma}} \, (1 + z)^{\eta} \, ,
  \label{eq:mergerrate}
\end{equation}
where $\xi$ is the dark matter merger mass ratio and $A = 0.0144$, $\xi_{0} = \xScientific{9.72}{-3}$, $\alpha = 0.133$, $\beta = -1.995$, $\gamma = 0.263$, $\eta = 0.0993$ are taken to the best-fitting parameters determined by \citet{Fakhouri2010}. 

Since we wish to verify the likelihood of early, major events, we integrate the merger rate between $z=5$ and $z=1$ along the median mass accretion history determined in Section~\ref{sec:sec:mergerscenarios}, first for mass ratios greater than 1:5. We obtain $N(0.2 \leq \xi \leq 1) = 2.01$, which is consistent with our merger scenarios in which Milky-Way-mass progenitors experience two significant events between $z=5$ and $z=1$ (Table~\ref{table:mergers}). This confirms that GSE-like feature are likely to be common across the Milky-Way analogue population (see also \citealt{Bignone2019, Mackereth2019, Fattahi2019, Elias2020, Dillamore2022}), and that our findings that GSE-like stars have several origins due to multiple high-redshift progenitors is likely generic (see also \citealt{Grand2020, Orkney2022GSEDouble}). 

The average merger rate in Equation~\ref{eq:mergerrate} strongly depends on the merger mass ratio $\xi$, rapidly suppressing increasingly major mergers (e.g. \citealt{Fakhouri2008, Fakhouri2010}). Re-integrating the merger rate with $\xi \geq$ 1:4, 1:3 and 1:2 gives $N = 1.67$, $N = 1.28$, and $N= 0.76$, respectively. Our genetically modified scenario with the most massive merger undergoes two events with dark matter mass ratios greater than 1:3 between $z=5$ and $z=1$, and is thus a less likely realization according to this metric. Interestingly, such rarer formation scenario could provide a natural mechanism to populate the small population of low-mass, bulge-dominated, central galaxies surrounded by low-surface brightness, star-forming rings (e.g. \citealt{Cappellari2011, Duc2015}). 

Despite this, we do not consider this last scenario (or any others) to be unlikely in a $\Lambda$CDM universe. Our genetic modifications generate minimal shifts in the likelihood of modified initial conditions compared to their reference, ensuring that all of them are highly consistent with the $\Lambda$CDM power spectrum (Section~\ref{sec:sec:mergerscenarios} and Figure~\ref{fig:massgrowths}). Furthermore, calculations of average merger rates suffer from systematic uncertainties due to issues in identifying haloes, trimming their merger trees and defining an infall mass ratio (e.g. \citealt{Fakhouri2008}). For example, using an updated merger tree algorithm and different snapshot cadence, \citet{Poole2017} report a much flatter dependence on the merger mass ratio for the same fitting function, particularly affecting the major merger rate. Recomputing the integral using their best-fitting parameters (their table 2), we find $N = 1.90$ for $\xi \geq$ 1:3, i.e. a 50 per cent increase making it compatible at face value with our most extreme scenario.

Future studies more robustly assessing the likelihood of a given merger history would thus be highly beneficial. This could be achieved by leveraging the extremely large halo catalogues of modern simulations (e.g. \citealt{Ishiyama2021, Maksimova2021}) to sample additional parameter space explicitly and construct joint likelihoods, or using more advanced statistical models than fitting functions to summarize the complex information content of halo merger trees (e.g. \citealt{Robles2022}). 

\section{Conclusion} \label{sec:conclusion}

We present a suite of genetically modified, cosmological zoomed simulations of Milky-Way-mass galaxies, systematically studying how varying the mass ratio of a $z\approx2$ merger impacts the $z=0$ chemodynamics of halo stars.

We start from a reference galaxy in a Milky-Way-mass dark matter halo ($\Mvir \approx 10^{12} \, \Msol$) which undergoes a large merger at $z\approx2$ ($\Mvir$ ratio of 1:6; $\vrvtheta = 16$). We then construct four alternative versions of this galaxy, systematically growing and decreasing the significance of this merger using the genetic modification approach (\citealt{Roth2016, Rey2018, Stopyra2021}). The suite scans stellar mass ratios of 1:24, 1:15, 1:8, 1:4, and 1:2 for the interaction, covering the range of inferred values for a potential GSE-progenitor when interpreting Galactic data (e.g. \citealt{Helmi2018, Kruijssen2020, Mackereth2020, Naidu2020, Naidu2021GSECharacterisation, Feuillet2021, Limberg2022}). 

All simulations are evolved to $z=0$ with the extensive hydrodynamical, galaxy formation model of the \textsc{vintergatan} project (\citealt{Agertz2020Vintergatan}), that can capture the multiphase structure of the galaxy's ISM ($\approx 20 \, \pc$) and a Hubble time of disc formation and dynamical evolution. Each genetically modified galaxy has, by construction, the same dynamical and stellar mass at $z=0$ (Figure~\ref{fig:massgrowths}) and a similar large-scale environment (Figure~\ref{fig:redshift3}). Comparing them with one another then provides a controlled, study isolating the signatures of a galaxy's early merger history on its chemodynamical structure at $z=0$. 
 
At fixed dynamical and stellar mass, modifications of the mass ratio at $z\approx2$ result in a large diversity of galactic structure and morphology at $z=0$ (Figure~\ref{fig:overview}). Smaller mergers favour the growth of an extended, rotationally supported stellar disc, while, by contrast, larger mergers leads to more compact, bulge-dominated morphologies after a Hubble time. In fact, our largest merger mass ratio at $z=2$ lacks a well-defined galactic disc, and exhibits a counterrotating inner core surrounded by a faint, blue star-forming disc similar to those observed around nearby ellipticals (e.g. \citealt{Duc2015}).

Despite the large diversity in galaxy morphology and structure, each galaxy with well-ordered stellar rotation exhibits a population of radially biased, inner halo stars overlapping in energy with disc stars (Figure~\ref{fig:phasespace} and Figure~\ref{fig:vrvphi}). We detect this kinematic GSE-like population with similar magnitude in all merger scenarios, whether we enhance the $z\approx2$ merger to become a major event or diminish it into a minor event, at odds with common interpretations linking GSE-like features to a single radially biased collision.

Furthermore, we show, for all merger mass ratios, stars within these GSE-like features have similar median peaks in ages (Figure~\ref{fig:radialstarsSFR}), $\feh$ (Figure~\ref{fig:radialstarsFeH}), and overlapping 2D chemodynamical distributions (Appendix~\ref{app:extradata}). This demonstrates that the existence and magnitude of a GSE-like population does not directly and causally relate to the mass ratio of a single, early merger at $z\approx2$, cautioning studies that aim to reconstruct the properties of an assumed single progenitor from $z=0$ chemodynamical data.

To identify the source of such degenerate signatures between multiple merger histories, we track back every GSE-like stars through the cosmological merger trees and pinpoint their origin. We find that halo stars on radial orbits at $z=0$ have diverse origins, originating from a mixture of \textit{in-situ} and \textit{ex-situ} subpopulations with overlapping chemodynamical distributions (Figure~\ref{fig:radialstarsFeH}; see also \citealt{Grand2020, Orkney2022GSEDouble}). 

In particular, we find that a significant fraction of the $z=0$ feature originates from stars formed during an earlier, high-redshift interaction at $z\approx3$, which have similar chemical abundances to the accreted population at $z\approx2$. As we decrease the significance of the progenitor at $z\approx2$ using genetic modifications, the contribution of its accreted component diminishes accordingly. However, this is compensated by the growth of the earlier, starburst population to conserve the same total stellar and dynamical mass, resulting in similar chemodynamics for the GSE-like feature.

Our study highlights the importance of modelling the full cosmological formation scenario when interpreting Galactic data at $z=0$. We explicitly demonstrate the importance of capturing the multiple high-redshift interactions, as well as the response of the central galaxy to each of them to obtain a full account of the several contributors to the final metal-poor, radial halo population. Robust inferences of past merger progenitors thus require a wide exploration of cosmological early scenarios, sampling a spectrum of progenitors, mass ratios, and infall geometry, to establish robust signatures distinguishing them. As shown by this study, the genetic modification approach combined with high-resolution zoomed simulations offers great potential to efficiently achieve this objective, and inform us on the cosmological formation scenario of our Milky Way. In particular, further genetic modifications controlling the angular momentum accretion history from the initial conditions (\citealt{Cadiou2021AngMomGM, Cadiou2022}) could allow us to vary the infall times and orbital parameters of GSE progenitors, in addition to their mass ratios studied here.

Despite the mechanisms and chemodynamical degeneracies between merger scenarios highlighted in this study, it remains unclear how to link one of our cosmological scenario (if any) to our Milky Way. All of our genetically modified galaxies have lower final stellar masses ($\Mstar \approx \xMsol{2}{10}$) than our Galaxy ($\geq \xMsol{6}{10})$, and blindly upscaling stellar masses across the merger tree would bring our GSE progenitors towards the massive end of commonly inferred values ($\Mstar \geq 10^9\, \Msol$), without changing their merger mass ratios. Our quieter genetically modified histories that prefer smaller dwarfs could then relate more closely to the potential history of our Milky Way, although a key finding from our study is that inferring past merger mass ratios from $z=0$ chemodynamical data is challenging. Furthermore, the numerical modelling of the stellar mass growth of galaxies, particularly that of dwarf galaxies at $z\approx2$, remains highly uncertain (e.g. \citealt{Somerville2015, Naab2017}). This adds further uncertainty and scatter to the dark matter halo masses that should host a given stellar-mass progenitor, and thus their dynamical impact on the protogalaxy. In addition to explorations of cosmological merger scenarios, future studies scanning through galaxy formation models will be key to understand how such uncertainties affect the reconstruction of our Milky Way's history.

Furthermore, our galaxies lack defining late-time features of our Galaxy (e.g. a bar), pointing to different evolution, gas accretion, and star formation activity at late times ($z<1$) between our scenarios and the Milky Way. Since early and late histories are highly correlated in a $\Lambda$CDM Universe, constraining the later evolution of our Galaxy will in turn reduce the available freedom in possible early, high-redshift history. The structure of the Milky Way's discs (e.g. \citealt{Belokurov2022Aurora, Xiang2022}; Figure~\ref{fig:overview}) and their potential accreted components (e.g. \citealt{Ruchti2015, Feuillet2022}), the density profile of the stellar halo (e.g. \citealt{Deason2011, Deason2014, Han2022StellarHaloProfile}), or the distribution of globular clusters (e.g. \citealt{Kruijssen2020}, although see \citealt{Pagnini2022}) offer us additional and complementary clues that can be combined with the GSE's properties to constrain the cosmological formation scenario of the Milky Way. 

The richness of Galactic data also offers numerous additional promising avenues to find more detailed diagnosis that would allow us to distinguish each merger scenario. Notably, using additional abundance ratios tracking different enrichment time-scales (e.g. \citealt{Belokurov2022Aurora, Feuillet2022,Matsuno2022SequioaAbundances}), combined with more detailed clustering techniques in the high-dimensional chemodynamical space to isolate subpopulations (e.g. \citealt{Myeong2022, Dodd2023}) will become ever-more powerful with the next-generation of spectroscopic surveys and the improvement in data quality and sample sizes in the coming years (e.g. WEAVE, 4MOST). 

\section*{Acknowledgements}
We would like to thank the referee for a constructive review that improved the quality of the manuscript. MR would like to thank Vasily Belokurov, Alyson Brooks, Amina Helmi, Kathryn Johnston, Harley Katz and Sarah Pearson for insightful discussions and comments during the construction of this work. MR is supported by the Beecroft Fellowship funded by Adrian Beecroft. OA and FR acknowledge support from the Knut and Alice Wallenberg Foundation, the Swedish Research Council (grant 2019-04659) and the Royal Physiographic Society in Lund. TS is supported by a CIERA Postdoctoral Fellowship. AP is supported by the Royal Society. This project has received funding from the European Union’s Horizon 2020 research and innovation programme under grant agreement No. 818085 GMGalaxies. We acknowledge PRACE for awarding us access to Joliot-Curie at GENCI/CEA, France to perform the simulations presented in this work. Computations presented in this work were in part performed on resources provided by the Swedish National Infrastructure for Computing (SNIC) at the Tetralith supercomputer, part of the National Supercomputer Centre, Link\"oping University. Development work was carried out on facilities supported by the Research Capital Investment Fund (RCIF) provided by UKRI and partially funded by the UCL Cosmoparticle Initiative.

We thank the developers of \textsc{genetic} (\citealt{Stopyra2021}), \textsc{jupyter} (\citealt{Ragan-Kelley2014}), \textsc{matplotlib} (\citealt{Hunter2007}), \textsc{numpy} (\citealt{vanderWalt2011}), \textsc{pynbody} (\citealt{Pontzen2013}), \textsc{scipy} (\citealt{Virtanen2020}), and \textsc{tangos} (\citealt{Pontzen2018}) for providing open-source softwares used extensively in this work. The Astrophysics Data Service (ADS) and arXiv preprint repository were used in this work.

\section*{Data availability}

The data underlying this article will be shared upon reasonable request to the corresponding author. The IllustrisTNG data used in this article is available at \url{https://www.tng-project.org}, and we will share derived data products upon reasonable requests.

\section*{Author contributions}
The main roles of the authors were, using the CRediT (Contribution Roles Taxonomy) system\footnote{\url{https://authorservices.wiley.com/author-resources/Journal-Authors/open-access/credit.html}}:

MR: Conceptualisation; Data curation; Formal analysis; Investigation; Writing – original draft. OA: Funding Acquisition; Investigation; Resources; Software; Writing – review and editing. TS: Conceptualisation; Writing – review and editing. FR: Writing – review and editing. GJ: Data Curation; AP: Conceptualisation; Writing – review and editing. NM: Writing – review and editing. DF: Writing – review and editing. JR: Writing – review and editing.

%%%%%%%%%%%%%%%%%%%%%%%%%%%%%%%%%%%%%%%%%%%%%%%%%%



\bibliographystyle{mnras}
\bibliography{GM_GES} 

%%%%%%%%%%%%%%%%%%%%%%%%%%%%%%%%%%%%%%%%%%%%%%%%%%

%%%%%%%%%%%%%%%%% APPENDICES %%%%%%%%%%%%%%%%%%%%%

\appendix

\section{Additional chemodynamical characterization of the GSE-like populations} \label{app:extradata}

\begin{figure*}
  \centering
    \includegraphics[width=\textwidth]{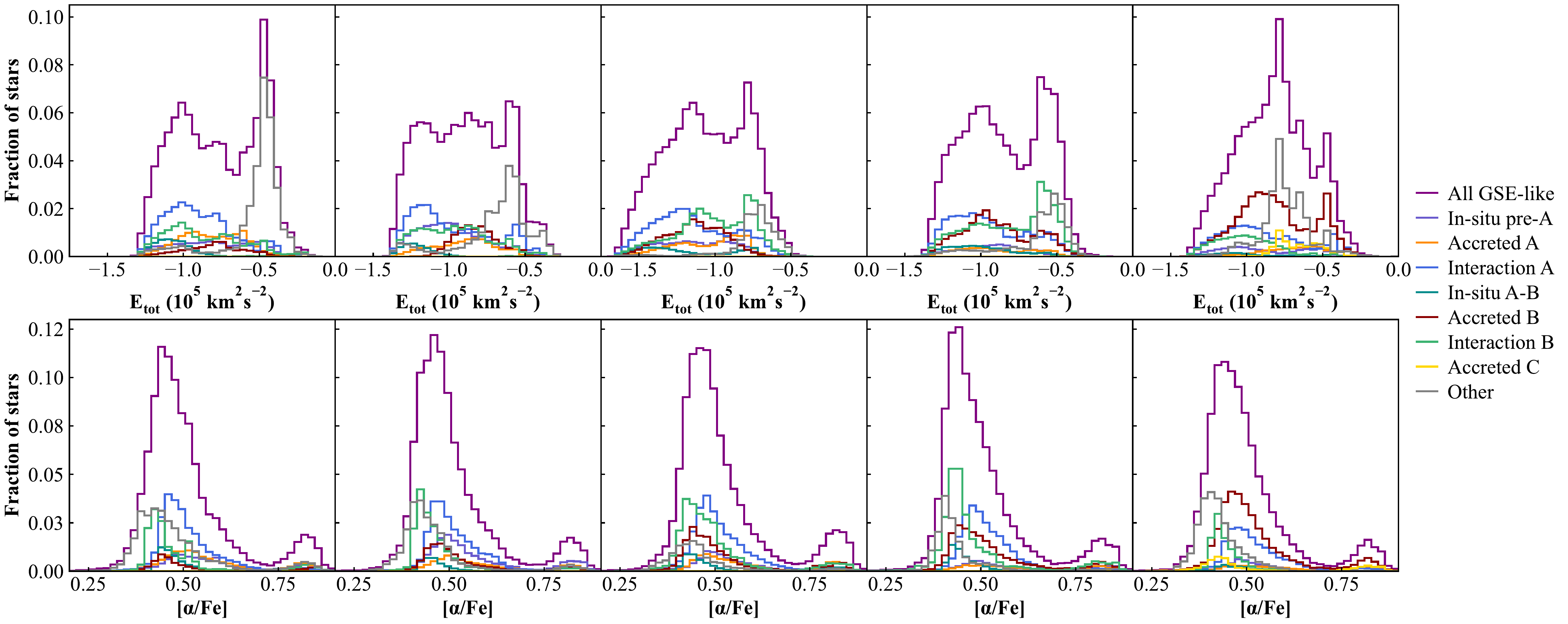}

    \caption{$\etot$ (top) and $\alphafe$ (bottom) distributions of all radially biased, GSE-like stars identified in Figure~\ref{fig:phasespace} (purple) and their high-redshift origin (colours matching Figure~\ref{fig:radialstarsFeH}). In all merger scenarios (increasing mass ratio from left to right), \textit{in-situ} and \textit{ex-situ} contributors substantially overlap chemically and kinematically, and cannot be easily distinguished (see also Figure~\ref{fig:debris2Dchemodynamics}). The contaminating population (grey) lies preferentially at high $\etot$ and low $\alphafe$.}
    \label{fig:debrischemodynamics}
\end{figure*}

\begin{figure*}
  \centering
    \includegraphics[width=\textwidth]{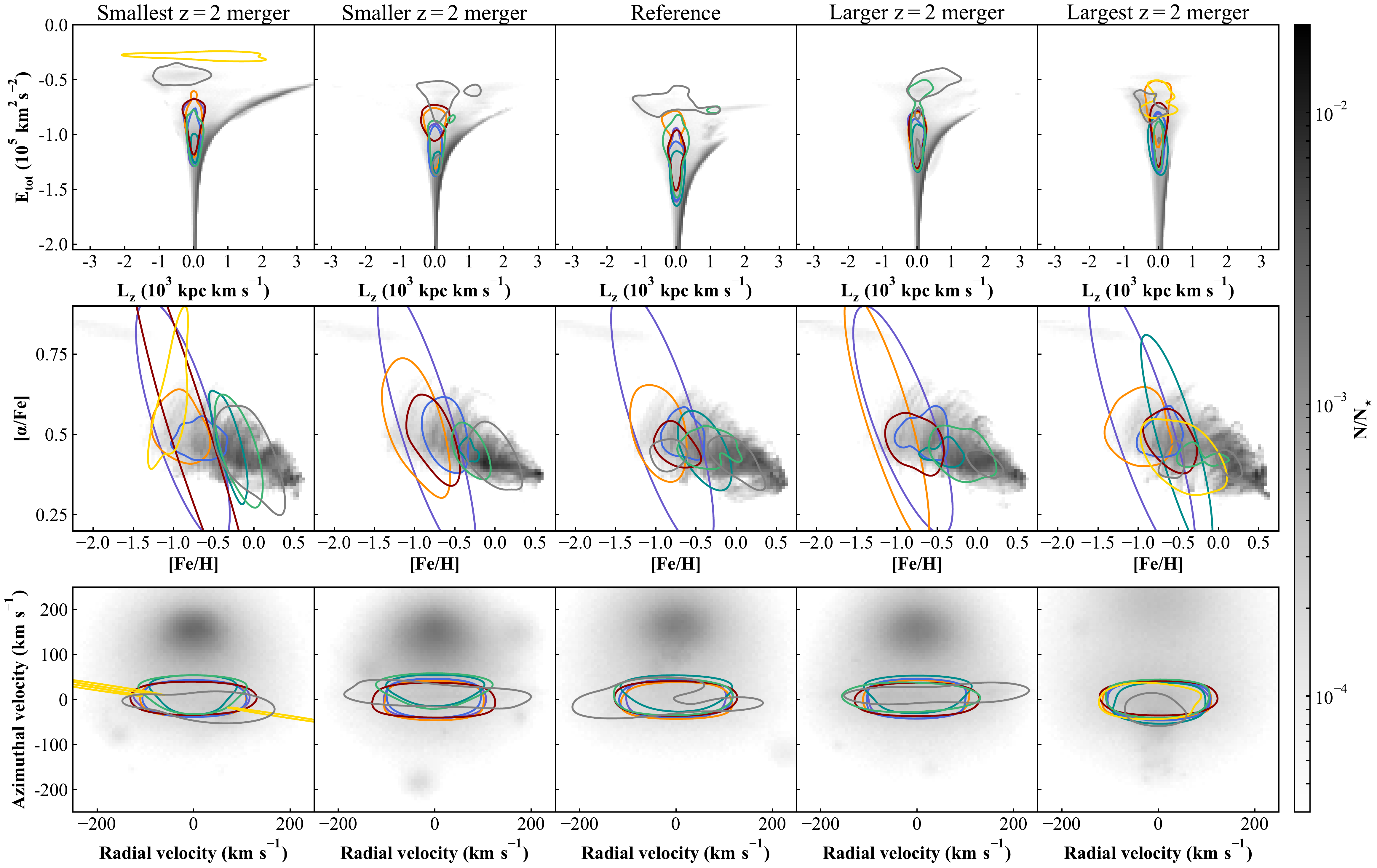}

    \caption{2D distributions in orbital (top), chemical (middle), and kinematic (bottom) space of each subpopulation identified in Figure~\ref{fig:radialstarsFeH}. Contours show 50\% isodensities in these planes, normalized to each subpopulation's total (their absolute contribution to the overall GSE-like feature can be estimated from Figure~\ref{fig:radialstarsFeH} and~\ref{fig:debrischemodynamics}). The most important subpopulations to the metal-poor GSE feature (blue and red) always overlap in all planes, making it difficult to distinguish increasing merger mass ratios at $z\approx2$ in the GSE-like progenitor (left to right). The grey histogram shows the distribution of all inner stars ($r < 50 \, \kpc$).
    }
    \label{fig:debris2Dchemodynamics}
\end{figure*}

In Section~\ref{sec:chemodynamics}, we establish that, despite a large diversity in final morphology and past merger mass ratio at $z\approx2$, each galaxy has similar GSE-like kinematic features at fixed stellar mass (Section~\ref{sec:chemodynamics:ges}). We further show in Section~\ref{sec:chemodynamics:originradialdebris} that GSE-like stars, in all merger scenarios at $z\approx2$, have broad age and $\feh$ distributions that peak around the same median, due to overlapping populations with varying cosmological origins. In this Appendix, we show that distinguishing subpopulations within each merger scenario, or merger scenarios from one another, remains difficult even when leveraging additional chemodynamical data for the GSE-like stars.

Figure~\ref{fig:debrischemodynamics} shows the distributions of $\etot$ and $\alphafe$ of each GSE subpopulation identified in Section~\ref{sec:chemodynamics:originradialdebris}. As in Section~\ref{sec:chemodynamics}, we recover strong similarities in the overall GSE-like population (purple) between the different merger scenarios, with similar peaks in abundances -- $\langle \alphafe \rangle = 0.48$, $0.48$, $0.48$, $0.47$, $0.47$, from left to right, respectively -- and broad coinciding distributions in $\etot$.

\begin{figure*}
  \centering
    \includegraphics[width=\textwidth]{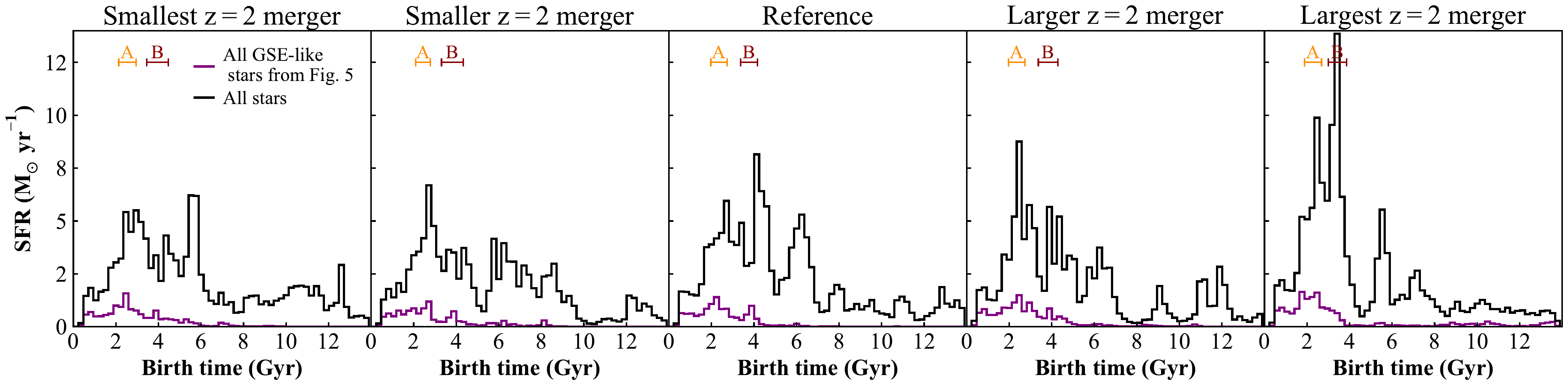}

    \caption{Same as Figure~\ref{fig:radialstarsSFR}, but comparing the galactic and GSE-like star formation histories. Peaks of star formation in GSE-like stars broadly map onto galaxy-wide star formation enhancements during early interactions (marked at the top). However, the opposite is not verified -- galaxy-wide starbursts do not necessarily form stars that end up on radial halo orbits at $z=0$. This is particularly visible for the largest mass ratio (right-most panel) where interaction `B' drives a large galaxy-wide starburst that is missing in the GSE-like population.}
    \label{fig:sfrcomparison}
\end{figure*}

Breaking down the overall population into its subpopulations (colours matching Figure~\ref{fig:radialstarsFeH}), we first confirm that, for all merger scenarios, the contaminating population (grey) has high $\etot$ and low $\alphafe$, combining with its higher $\feh$. This is consistent with our interpretation that this contamination primarily arises from outer disc and halo stars that formed after the ancient structures of interest. 

Focusing on the subpopulations originating at high-redshift (colours), we recover a strong overlap in $\etot$ and $\alphafe$, adding to the kinematic and chemical overlap already highlighted in Section~\ref{sec:chemodynamics}. In particular, stars formed during early interactions (light blue and green) can populate high-$\etot$ orbits and have similar $\alphafe$ ratios than the accreted population of the genetically modified merger at $z\approx2$ (`B'; red).

To further visualize this, we show in Figure~\ref{fig:debris2Dchemodynamics} the positions of subpopulations in the $\etot$-$\jz$, $\alphafe$-$\feh$, and $\vr-\vphi$ planes (top, middle and bottom, respectively) compared to the overall population of stars within the galaxy (grey background). We estimate respective 50\% isodensity contour using a 2D kernel density estimate with Gaussian smoothing according to Scott's rule. The small sample sizes, large spread and non-Gaussian nature of chemodynamical distributions can lead to noisy KDE estimates\footnote{We find that going for larger isodensity contours (e.g. 68\% for one sigma) exacerbates this issue.}, and generates the large isodensity contours for very early populations that have large spreads in $\alphafe$ and small sample sizes. For completeness, we choose to show all subpopulations, even when noisy, but stress that they might not contribute a large fraction of the total GSE-like feature (absolute contributions can be estimated from Figure~\ref{fig:radialstarsFeH} and~\ref{fig:debrischemodynamics}). 

As with 1D distributions, we find substantial similarities between merger scenarios (left to right). Individual subpopulations have distinct peaks and a clear progressing sequence in $\alphafe$-$\feh$. However, the main contributors to the metal-poor GSE-like population (blue formed during the earlier interaction `A', and red brought with the merger body `B') all significantly overlap in these planes. We thus do not find differences in chemodynamics of the GSE-like population that unequivocally allows us to differentiate merger mass ratio at $z\approx2$.

\section{Comparing galaxy-wide and GSE star formation histories} \label{app:starburst}

In Section~\ref{sec:chemodynamics:originradialdebris}, we show that star formation peaks in the GSE-like population coincide in timing with the coalescence of early mergers (Figure~\ref{fig:radialstarsSFR}), when a significant fraction of its stars are formed (Figure~\ref{fig:radialstarsFeH}). In this Appendix, we quantify whether this surplus in star formation reflect a general galaxy-wide enhancement, or is specific to stars on halo-like, radial orbits at $z=0$.

We show in Figure~\ref{fig:sfrcomparison} the star formation histories of all inner stars ($r \leq 50 \, \kpc$; black) for each merger scenario (mass ratio at $z\approx2$ increasing from left to right), compared to those of GSE-like stars (purple; see also Figure~\ref{fig:radialstarsSFR}). We recover that early interactions ($\zinfall$ to $\zend$ marked at the top) drives large, galaxy-wide star formation enhancements consistent with merger induced starbursts. For lower mass ratios at $z\approx2$ (first three panels), the elevated star formation coincide with peaks in star formation in the GSE-like population, reflecting the dominant contribution of starburst stars in this population at $z=0$ (Figure~\ref{fig:radialstarsFeH}).

For larger mass ratios (right-hand panels), the mapping between galaxy-wide and GSE star formation is less clear, particularly for interaction `B' (red). This is particularly visible for the largest mass ratio (right-most panel), where the galaxy undergoes a large starburst during interaction `B' (black), but stars formed at this time do not populate radially biased orbits at $z=0$ (purple). In fact, the origin of GSE-like at $z=0$ population is dominated by accreted stars previously formed in the merging body, rather than stars formed during the interaction (Figure~\ref{fig:radialstarsFeH}).

%%%%%%%%%%%%%%%%%%%%%%%%%%%%%%%%%%%%%%%%%%%%%%%%%

% Don't change these lines
\bsp	% typesetting comment
\label{lastpage}
\end{document}